\documentclass[12pt,a4paper,oneside]{article}			\usepackage{subcaption}
\usepackage{caption}
\usepackage[margin=2.5cm]{geometry}	 						
\usepackage{setspace,booktabs,lscape,soul,dirtytalk}
\rmfamily												 			 				       \usepackage[utf8]{inputenc}					\usepackage{color}										 \definecolor{ahjcolor}{rgb}{0.0, 0.13, 0.28} 			
\usepackage[colorlinks	=true,							            linkcolor	=ahjcolor,						  urlcolor	=ahjcolor,
            citecolor	=ahjcolor]{hyperref}					\usepackage{multicol}
		
\usepackage{amsmath}	
\usepackage{bm}	
\usepackage{amsfonts,threeparttable}
\usepackage{amssymb}
\usepackage{graphicx}
\usepackage{multirow}
\usepackage{hyperref}	\usepackage{natbib}
\usepackage{caption}	
\usepackage{wrapfig}
\usepackage{sectsty}									
\singlespace											 \chapterfont{\centering \sc \Large}						 \sectionfont{\centering \sc \large}						 \subsectionfont{\sc \large}								 \subsubsectionfont{\sc \large}							 \usepackage{fancyhdr}										 										 												 
\usepackage{filecontents}

\usepackage{floatrow}   \title{Indirect Estimators of Intergenerational Mobility\thanks{\textbf{Preliminary version} in preparation for the \textit{Handbook of the Economics of Intergenerational Mobility}}}

\author{Andrea Del Pizzo\thanks{Department of Economics, Universidad Carlos III de Madrid, Spain. Email: \href{mailto:apizzo@eco.uc3m.es}{apizzo@eco.uc3m.es}.}
\and
Martin Nybom\thanks{IFAU and Uppsala University, Sweden. Email: \href{mailto:martin.nybom@ifau.uu.se}{martin.nybom@ifau.uu.se}.}
\and
Jan Stuhler\thanks{Department of Economics, Universidad Carlos III de Madrid, Spain. Email: \href{mailto:jstuhler@eco.uc3m.es}{jstuhler@eco.uc3m.es}.}}

\begin{document}

\maketitle
\begin{abstract}
    This chapter reviews indirect estimators of intergenerational mobility, focusing on approaches that infer parent–child or other family associations when direct income data are incomplete or unavailable. We synthesize methods based on instrumental variables, imputation using observable characteristics such as education and occupation, surname-based estimators, and multigenerational linkages. To unify these approaches, we introduce a stylized framework in which socioeconomic status is transmitted through multiple pathways with heterogeneous persistence rates. Within this framework, both direct and indirect estimators can be interpreted as weighted averages of these underlying transmission channels. A central insight is that the choice of instrument or imputation strategy determines these weights, leading different methods to capture distinct aspects of the transmission process. We highlight implications for interpretation, showing that indirect estimators need not recover conventional parent–child correlations but can instead provide complementary evidence on long-run persistence and the mechanisms underlying persistent inequalities.
\end{abstract}

\thispagestyle{empty}
\setcounter{page}{0}



\newpage
\section{Introduction}\label{sec:intro}

To assess the strength of intergenerational transmission, researchers typically relate parents’ socioeconomic status to that of their children. But rather than directly linking parent to child outcomes, a growing literature adopts \textit{indirect} methods. These approaches are often motivated by data limitations, but increasingly also by the aim to gain deeper insights into mobility processes that may not be apparent from direct estimates alone. A common approach is to impute parental income using other parental characteristics, such as education or occupation, in an instrumental variable (IV) design. More recently, some studies have used \textit{names} as proxies for parental status, or have drawn on more distant relatives, either to instrument parental status or to study how socioeconomic correlations vary across kinship distance.

This chapter reviews such indirect estimators of intergenerational mobility. We describe each approach and highlight key studies that have developed or applied them, though our primary focus is on how these estimators should be \textit{interpreted}. To guide our discussion, we introduce a simple statistical framework capturing two essential features of transmission processes: (i) socioeconomic status is transmitted through multiple pathways or mediators (such as 
skills, inheritances, neighborhood characteristics, or social networks), and (ii) the rate of transmission may vary across pathways. Although simple, these assumptions have fundamental implications for the interpretation of mobility estimates.

Our key argument is that different estimators should be interpreted as 
differently weighted combinations of the underlying transmission pathways. This argument applies to both direct and indirect estimators of intergenerational mobility. For indirect estimators, however, the weights depend crucially on the choice of instrument and can be heavily concentrated on certain pathways. Name-based estimators place greater weight on pathways with high persistence, especially when applied to larger name groups. A similar argument can be made for multigenerational approaches that use the status of more distant family members as instruments. We illustrate these points using linked U.S. Census data and register data from Sweden.

Before describing the different estimators, Section \ref{sec:model} introduces a stylized framework that allows for multiple, potentially latent, transmission channels with heterogeneous rates of persistence (``\textit{multiplicity}''). The simple latent factor model, which motivated some recent name-based and multigenerational studies, is a special case of our framework. 
 
Section \ref{sec:IV} reviews conventional IV approaches, used in situations in which the standard parent-child correlation is the explicit target parameter, but the information on parental earnings or income is incomplete. When parental income is poorly measured, measurement error can be corrected for using other determinants of income, such as education or occupation, as instruments \citep{zimmerman1992regression,solon1992intergenerational}. When parental income is missing entirely in the linked sample, a two-sample approach can be used to predict parental income with the help of an auxiliary dataset \citep{bjorklund1997intergenerational}. As comparative cross-country studies rely heavily on this two-sample approach (\citealt{narayan2018fair, munoz2025intergenerational}), particularly in settings such as developing countries where direct estimation is often not feasible, it is important to understand what these estimators identify and how they relate to direct estimates.

As we argue, the estimands identified by such IV approaches depend critically on the choice of instrument—specifically, on how instruments correlate with different pathways of transmission and on the persistence of those pathways. We confirm this argument in Swedish register data, where we can compare direct estimates of income persistence with IV estimates based on different instruments. Among other things, this helps explain the common finding that indirect IV estimates exceed direct estimates:  observable characteristics typically used as instruments, such as education, tend to be more persistent than unobservable determinants of income, such as market luck. Yet, even across typically available observables, IV estimates vary widely with the choice of instrument and its persistence. In our data, for example, the IV estimate is twice as large as the direct estimate when using region of residence but only 25\% higher when using occupation. These findings are consistent with recent work showing that appropriate instrument choice can yield estimates closer to the direct estimate (\citealt{bloise2021estimating}).

Section \ref{sec:names} examines name-based approaches, where researchers use surnames—or, in some cases, first names (\citealt{olivetti2015name})—to predict individuals' socioeconomic status \citep{clark2014also,guell2015informational}. While methods vary, the most common one involves estimating persistence at the surname level, an approach that can be interpreted as an IV estimator, where surnames serve as predictors for parental income or status. Our argument that IV estimators weight transmission pathways differently extends to this context. As shown in \cite{delpizzo2025in}, name-based estimators implicitly place greater weight on more persistent transmission channels, as surnames capture similarities between \textit{distant} relatives, which are driven by highly persistent traits. Such traits may include those related to geography (see \citealt{Clarketal2015Surnames, guell2018correlating, longley2021geography}) or race (\citealt{mazumder2021race, jacome2025mobility,ward2025like}). 
 
Our framework thus delivers a simple explanation for two common findings in this literature: name-based estimates can be much larger than direct estimates (\citealt{clark2014also}), and tend to increase in the size of surname groups (\citealt{chetty2014land}). We document both patterns in U.S. Census data \citep{ruggles2024ipums}. However, sampling properties also matter, especially to understand why name-based estimates differ so much across studies (\citealt{SantarvitaStuhlerEJ}). A key distinction lies in whether surname averages for parents and children are constructed from the same families (\textit{overlapping samples}, as in full-count census data) or from different families (e.g., when using subsamples from the census). The logic can be understood by drawing on the weak instruments literature: overlapping samples resemble 2SLS, with estimates biased towards OLS, while in independent samples the estimates are biased towards zero. This distinction is especially relevant in name-based research, as name instruments are weak but numerous.

Section \ref{sec:multigenerational} reviews ``extended-kin'' or ``dynastic'' approaches to intergenerational mobility. While the previous two strands emerged in response to data limitations, this line of research is driven by growing data availability that allows researchers to link family members across more than two generations, to deepen our understanding of mobility processes. A fast-growing body of \textit{multi}generational evidence shows that simple extrapolations from conventional two-generation measures tend to understate the extent of status persistence across multiple generations \citep{lindahl2015long, braun2018transmission}. Recent work extends this literature by linking across more generations \citep{hallsten2023shadow, shiue2025social}, or by incorporating assortative mating and horizontal family ties \citep{adermon2021dynastic, collado2023estimating, clark2023inheritance, bingley2023origins, FamilyCookbook}. 

While this dynastic information is used in different ways, the key idea is that a comparison of close and distant kin can be indirectly informative about properties of the transmission process that are not observable from parent-child correlations alone. In this sense, the extended-kin approach shares important commonalities with the indirect IV and name-based approaches reviewed in earlier sections. In particular, we argue that each approach should be viewed as producing differently weighted combinations of the underlying transmission mechanisms. The key distinction lies in motivation: in conventional IV and name-based studies, such reweighting typically arises as an unintended byproduct, whereas in multigenerational and dynastic settings it is more often a deliberate and explicit feature of the research design.

We conclude by highlighting a set of implications. First, for all indirect estimators, the choice of instrument, or mode of imputation, is crucial. Some instruments place greater weight on highly persistent transmission pathways, others on less persistent channels. As we argue, there is a logic to why different instruments used in the literature can yield quite different estimates. Second, indirect approaches are unlikely to recover the parent-child income correlation. If such direct mobility measures are indeed the target parameter, certain instruments may be more appropriate than others.

For some questions, however, indirect estimators can be \textit{more} informative than direct ones. For example, surname-based estimators may be particularly informative about the rate at which socioeconomic differences decay in the very long run. Moreover, the interpretive challenges that apply to indirect estimators also affect direct ones, as conventional parent-child correlations too reflect only a particular weighted combination of underlying transmission mechanisms. While appropriate for describing the similarity of parents and children in a given outcome such as income, direct estimates may provide limited insight into the strength of underlying transmission processes, broader socioeconomic persistence, or the dynamics of income advantages beyond two generations. The indirect methods reviewed in this chapter can thus offer additional insights into how status advantages are transmitted across generations.

\section{Conceptual Framework}\label{sec:model}

We use a stylized model of intergenerational transmission to shed light on the interpretation of various indirect estimators from the literature. While our arguments could be applied to other socioeconomic characteristics, such as education or occupation, we use income as our running example throughout the chapter. 

First, assume the income of individual $i$ in generation $t$, denoted $y_{it}$, depends on a set of underlying characteristics $X_{it}^{j}$, such as education, different types of abilities, and geographical location,
\begin{equation}
\label{intra_eq}
y_{it}=\beta_{0}+\sum_{j=0}^{J}\rho_{j}X_{it}^{j}+u_{it}
\end{equation}
In principle, $X_{it}^{j}$ will represent any characteristic that influences income (or whatever outcome is considered), whether observed (e.g., location) or unobserved (e.g., latent ability), and where the error term $u_{it}$ represents random deviations from the expected income based on these characteristics.

Second, each characteristic $X_{it}^{j}$ is transmitted from parents to children at a specific rate $\lambda_j$, which may differ across characteristics,
\begin{equation}
\label{inter_eq}
X_{it}^{j}=\lambda_{j}X_{it-1}^{j}+\epsilon_{it}^{j}
\end{equation}
where $\epsilon_{it}^{j}$ captures stochastic deviations in the relation between child and parent characteristics. The characteristics $X_{it}^{j}$, together with their returns $\rho_{j}$ and transmission rates $\lambda_j$, thus constitute the \emph{pathways} or \emph{channels} through which intergenerational persistence in a given outcome (e.g., income) arises.

Obviously, this model is stylized and restrictive in various respects. For example, equation (\ref{intra_eq}) does not allow for a direct causal effect of parent on child income, and equation (\ref{inter_eq}) assumes that each child characteristic depends only on the corresponding parent characteristic, while one would expect important cross-effects (e.g., parents' income or their choice of location may shape their child's educational attainment, \citealt{durlauf1996theory,HHBlandenDoepkeStuhler23}). However, plausible extensions such as mediating or cross-effects can be easily integrated and would have few implications for our core arguments.\footnote{For example, the parameter $\rho_{j}$ in eq. (\ref{intra_eq}) can be defined to include not only the direct effect of characteristic $X_{it}^j$ but also its indirect effects that operate through potential mediators. Similarly, the transmission rate $\lambda_j$ in eq. (\ref{inter_eq}) can be thought of as a reduced-form parameter that also reflects the strength of assortative mating. See \cite{collado2023estimating} for a latent factor model that accounts for direct effects, assortative mating, and shared sibling influences.} These arguments rest on two key assumptions, which we refer to as ``multiplicity'' (\citealt{stuhler2012mobility}): (i) intergenerational transmission operates through multiple pathways and (ii) these pathways differ in their rates of transmission.\footnote{The idea that intergenerational transmission occurs via multiple pathways is also emphasized in standard economic models, such as \cite{becker1979equilibrium}. While Becker and Tomes summarize these pathways in a one-dimensional ``endowment'' with a single persistence parameter, we emphasize that different pathways may exhibit distinct transmission rates (see also, e.g., \citealt{BlandenGreggMacmillan2007,bjorklund2017contribution}), and that such heterogeneity has important implications for the interpretation of intergenerational estimators. See Section \ref{sec:TSLS} for supporting evidence.}

The need for such a model, rather than simpler processes linking parental income to child income, such as an AR(1), stems from a well-established empirical regularity: the rate at which kinship correlations decay slows with genealogical distance \citep{lindahl2015long}. Let $\beta_g$ denote the intergenerational correlation at genealogical distance $g$. Empirically, iterating the parent-child $\beta_1$ forward systematically understates persistence across multiple generations, in the sense that $\beta_g > \beta_1^g$ for all $g>1$ (and assuming for simplicity that the parent-child correlation is stable across generations). This pattern motivates a distinction between \textit{short-run} persistence, summarized by $\beta_1$, and \textit{long-run} persistence, captured by $\beta_g$ for $g>1$.

This gap between short- and long-run persistence can be rationalized using a latent factor model: if observed outcomes are imperfect proxies of an underlying persistent endowment or status component \citep{clark2014also,braun2018transmission}, then parent-child correlations are attenuated measures of the persistence of the latent trait driving long-run outcomes.\footnote{The latent-factor model is not the only possible interpretation to explain why persistence increases with genealogical distance. Other potential interpretations include non-Markovian and non-linear transmission processes (\citealt{stuhler2024multigenerational, blume2025immobility}). In particular, a strand of the literature has explored whether grandparents have an independent effect on their grandchildren (``\textit{grandparent effects}'', \citealt{chan2013grandparents, HertelSamberg2013, long2018grandfathers}).} This single-factor latent model is a special case of the framework discussed above, corresponding to the restriction
\begin{equation}
\label{rest}
\lambda_{j}=\lambda_{k}=\lambda \hspace{20pt} \forall j\neq k .
\end{equation}
under which all underlying characteristics share a common transmission rate.

Both the single-factor model and our multiplicity framework imply that knowledge of the parent-child correlation ($\beta_1$) offers only limited insight about long-run persistence ($\beta_g$). However, they differ in how $\beta_g$ evolves with genealogical distance $g$. In a single-factor setting, the \textit{rate of decay} defined as $\beta_g / \beta_{g-1}$ is constant for $g\geq2$ and equal to $\lambda$, so that a single parameter fully determines the evolution of $\beta_g$ beyond $\beta_1$ (as $\beta_g = \prod_{k=1}^{g} \frac{\beta_k}{\beta_{k-1}}$). This parameter is thus key to understand long-run persistence. It might also be the key parameter to understand parent-child mobility, depending on how we interpret the structural model and the outcome variable. For example, if the observed outcome is a noisy proxy for the actual outcome of interest, $\beta_1$ is attenuated by measurement error and $\lambda$ may be a more meaningful measure of parent-child mobility.\footnote{For illustration, suppose observed income depends on two orthogonal components: a persistent component \textit{A}, which explains 50\% of the variance and is perfectly transmitted across generations, and a transitory component \textit{B}, which explains the remaining 50\% and is not transmitted. Then the intergenerational correlation equals 0.5 at all genealogical distances ($\beta_1 = \beta_g = 0.5$), while the rate of decay equals one. Whether the correlations or their rate of decay are a more meaningful measure of mobility depends on structural assumptions. If the transitory component reflects measurement error, $\beta_g$ overstates income mobility while the rate of decay captures ``true'' persistence; if instead it reflects market luck, $\beta_g$ accurately captures income persistence while the decay rate provides complementary information on long-run dynamics.} 

By contrast, under multiplicity, long-run persistence is no longer governed by a single parameter, as the decay rate $\beta_g / \beta_{g-1}$ varies with genealogical distance. While formally derived below, the intuition is straightforward. When characteristics differ in how strongly they persist across generations, weakly transmitted traits fade rapidly, while more persistent traits decline much more slowly. As genealogical distance increases, intergenerational correlations therefore become increasingly driven by highly persistent characteristics, even if these account for only a modest share of parent-child transmission. These compositional shifts imply a progressive slowing of convergence to the mean, rather than decay at a constant rate. 

As a result, in a model with multiplicity, no single parameter can summarize how persistence evolves in the long run. Instead, the rate at which kinship correlations decay is inherently distance-dependent, and this has important implications for how persistence can be summarized and estimated. Direct estimators linking parents to their children provide only limited information about multigenerational dynamics in $\beta_g$. Conversely, estimators capturing decay at higher genealogical distance $g$ may overstate the rate of decay at shorter distances. A comprehensive understanding of multigenerational dynamics therefore requires evidence from multiple empirical designs, as different approaches capture different aspects of persistence—an idea that will be central to what follows.

\subsection{Direct Intergenerational Estimators under Multiplicity}

What do intergenerational estimators capture if transmission operates through multiple and heterogeneous transmission channels? While our framework can be applied to intergenerational persistence in any outcome, we keep with our main example of income persistence. We begin by considering the conventional (``direct'') intergenerational income estimate, as is estimable if incomes are observed and linked for two generations:
\begin{align}\label{eq:corr2g}
    \beta_{\textit{direct}}=\frac{Cov(y_{it};y_{it-1})}{V(y_{it-1})}=\frac{\sum_{j=0}^{J}\lambda_{j}\omega_{j}}{\sum_{j=0}^{J}\omega_{j}+\omega_{u}}
\end{align}
where 
\begin{align}\label{eq:corr2gweights}
    \omega_{j}=\rho_{j}^{2}V(X_{it-1}^{j})+\rho_{j}\sum_{k\neq j}^{J}\rho_{k}Cov(X_{it-1}^{j};X_{it-1}^{k})
\end{align}
Viewed through our framework, the parent-child income correlation reflects a weighted average of the persistence rates $\lambda_j$ of all relevant factors affecting one's income. The weights $\omega_{j}$ are determined by each factor’s contribution to the variance of parental income, accounting for both its own effect and its covariance with other characteristics. 

Implicitly, this framework allows intergenerational mobility to differ across families or groups, as differences in the composition—and thus persistence—of underlying characteristics mechanically generate heterogeneous mobility rates.\footnote{The persistence of group-level differences has been emphasized in both economics (e.g., \citealt{borjas1992ethnic, Mazumder2014BlackWhite, solon2018we, chetty2020race, ward2020not, mazumder2021race,boustan2025intergenerational}) and sociology (\citealt{torche2018estimating, karlson2025decomposing}). A related literature considers heterogeneity in transmission rates across families (\citealt{bingley2019correlation,Puerta2025Family,chang2025accounting, HorlHetMobility}) or across the socioeconomic distribution (\citealt{colagrossi2024heterogenous, hjorth2025different}).} While deliberately simple, this decomposition provides a useful benchmark for the indirect estimators analyzed in the next sections.

\subsection{Indirect Estimators under Multiplicity}
\label{ind_estimator}
In many settings, a lack of direct parental income data leads researchers to rely on indirect methods to measure intergenerational persistence. We explore the identification properties of these indirect approaches and show how different types of imputation (or ``instruments'') yield different estimates of intergenerational income persistence. Our arguments mirror the broader insight that when treatment effects vary, different instruments identify different local average treatment effects, or LATEs \citep{imbens1994identification}. Indeed, most of the indirect estimators we discuss are explicit or implicit variations of standard instrumental variables techniques, and can be analyzed within the two-stage least squares (TSLS) framework.

As shown in eq. \eqref{eq:corr2g}, when intergenerational transmission operates through multiple pathways, the parent-child correlation $\beta_{\textit{direct}}$ captures a weighted average of the persistence rate of each pathway. However, when indirect methods such as IV-based imputation are used, the weighting structure adjusts based on the chosen instrument. Consequently, different instruments capture distinct transmission channels, leading to variation in estimated persistence, similar to the logic of the LATE theorem. 

For instance, consider a source of variation in parental income that is orthogonal to other forms of human capital, such as a lottery win. The evidence suggests that such shocks have modest effects on children’s long-term outcomes.\footnote{Lottery wins appear to have limited intergenerational effects, while the literature on cash transfers documents more positive impacts; see the chapter by \cite{nybomsocial} in this volume.} By contrast, increases in parental income driven by factors like educational attainment or cognitive and noncognitive skills are more likely to produce enduring effects across generations, operating through multiple channels. More generally, different mechanisms contribute to intergenerational income transmission, each with varying degrees of persistence. As a result, any given instrument identifies the persistence associated with the specific mechanisms it captures.
\par 
To formalize this intuition, we derive the model-implied formula for the TSLS estimator when applied to a general instrument $Z_{it-1}$ used to impute $y_{it-1}$ (e.g., income) in the parental generation. In a standard TSLS framework, the intergenerational estimator can be written as:\begin{align}
\label{beta_iv}
\beta_{\textit{TSLS}}=\frac{Cov(y_{it};E[y_{it-1}|Z_{it-1}])}{V(E[y_{it-1}|Z_{it-1}])}=\frac{\sum_{j=0}^{J}\lambda_{j}\omega_{j}^{Z}}{\sum_{j=0}^{J}\omega_{j}^{Z}}
\end{align}
where each weight takes the following form:
\begin{align}
    \omega^{Z}_{j}=\rho_{j}^{2}V(E[X_{it-1}^{j}|Z_{it-1}])+\rho_{j}\sum_{k\neq j}^{J}\rho_{k}Cov(E[X_{it-1}^{j}|Z_{it-1}],E[X_{it-1}^{k}|Z_{it-1}])
\end{align}
Ignoring cross-factor covariances for the moment (i.e. the second term above), the implicit TSLS weighting (the first term) consists of (i) an instrument-invariant component and (ii) an instrument-specific component. The first component, $\rho_{j}^{2}$, reflects the relevance of each characteristic. For example, if education is the primary determinant of income, its associated coefficient $\rho_{edu}$ will be large and its contribution will dominate those of other factors. This parameter is fixed by the data-generating process and does not depend on the choice of instrument.
\par 
The second component, $\operatorname{Var}\!\big(\,\mathbb{E}[X_{it-1}^j \mid Z_{it-1}]\,\big)$, depends instead on the instrument, measuring how strongly the instrument shifts factor \(j\) (i.e., its predictive content for \(X^j\)).
For example, if \(Z\) is parental education, then $\operatorname{Var}\!\big(\,\mathbb{E}[X^{\text{edu}} \mid Z^{\text{edu}}]\,\big) = \operatorname{Var}\!\big(X^{\text{edu}}\,\big)$, so the weight \(\omega^{\,\text{edu}}_{\text{edu}}\) is large. By contrast, if \(Z\) refers to geography and education
does not vary systematically across regions, then $
\operatorname{Var}\!\big(\mathbb{E}[X^{\text{edu}}\mid Z^{\text{geo}}]\big)=0
\quad\text{and}\quad
\omega^{\,\text{geo}}_{\text{edu}}=0.
$ In short, TSLS identifies persistence along the dimensions the instrument actually moves (e.g.,
human capital under an education instrument), not along dimensions it leaves unchanged.

This mechanism explains why the TSLS estimator is bound to differ with the choice of instruments: the more closely a given income-determining characteristic is correlated with the chosen instrument, the more its persistence rate contributes to the estimated parent-child correlation. While this observation implies that indirect methods are not necessarily resulting in accurate approximations of such correlation, they may capture properties of the transmission process that are interesting in their own right. For example, in Section \ref{sec:names} we argue that name-based estimators might be particularly informative about \textit{long-run} persistence. 

\section{IV Approaches with Incomplete Income Data}
\label{sec:IV}

In many settings, estimating direct measures of income mobility, such as the IGE, is complicated by missing or incomplete information on either parental or child income (for recent reviews, see  \citealt{nybom2024intergenerational,deutschermazumder2023measuring}).\footnote{Among earlier reviews of the intergenerational mobility literature, see \citep{solon1999intergenerational, black2011recent}.} To address such limitations, researchers have developed alternative approaches. When parental income is poorly measured (relative to lifetime incomes), a common strategy to correct for errors-in-variables bias is to instrument noisy income measures with more stable determinants of lifetime income, such as education or occupation \citep{zimmerman1992regression, solon1992intergenerational}. In other cases, the intergenerational sample (``main sample'') may lack information on parental income altogether. A popular workaround in such contexts is to use \emph{two-sample} two-stage least squares (TSTSLS), where parental characteristics, such as occupation, education, or even surnames (see next section), are used as instruments to predict parental income with the help of an \emph{auxiliary} dataset \citep{bjorklund1997intergenerational}.\footnote{A related approach is sometimes used in the literature on the assimilation of immigrants over generations, relying on a grouping estimation strategy using the mean earnings levels of immigrants, e.g., by cohort and country of origin, rather than intergenerationally linked individual data \citep{borjas1993intergenerational,card2000more}. This approach also shares similarities with grouping approaches based on name groups; see Section 4.}

This section reviews such ``conventional'' IV strategies for incomplete or missing parental income data. Our focus here is on the long-standing approach of using Mincer-style \emph{parental} characteristics, such as education or occupation, as instruments, an approach popularized in the 1990s. We first address the identification of the estimators within our transmission framework. Our arguments apply equally to both TSLS and two-sample TSLS (TSTSLS), and they are hence discussed jointly. We then discuss some methodological concerns that are specific to the two-sample case. Lastly, we review classic and recent 
applications, with particular emphasis on two-sample implementations. Name-based studies can be interpreted as a special case of the same IV approach, and we treat them separately in the next section.

\subsection{The Interpretation of Conventional TSLS Estimators}

Prior research has emphasized that TSLS yields biased estimates of the true parent-child correlation $\beta_{\textit{direct}}$, due to the non-excludability of commonly used instruments (\citealt{solon1992intergenerational,dearden1997intergenerational}). Since instruments like parental education or occupation likely have positive independent impacts on children, TSLS estimates have generally been seen as providing an upper bound for the direct estimate. Empirical evidence has supported this view. 
We revisit this argument by interpreting TSLS estimation within our framework. Our key observation is that different types of IV estimators, as formalized in Section \ref{ind_estimator}, can be understood as identifying different intergenerational parameters, characterized by alternative weighting structures. 

Specifically, the difference between the TSLS estimate and the direct parent-child estimator arises from two main factors. First, the IV approach abstracts from random influences on parental income, such as those arising from market luck or measurement error (e.g., \citealt{mazumder2016estimating}), because the conditional expectation of these random influences, given the instrument, is zero. Formally, $\omega_u^Z$ is absent in the denominator of eq. (\ref{beta_iv}) because $V(E[u_{it-1}|Z_{it-1}])=0$, leading to an upward adjustment of the estimate. This feature was noted already by \citet{solon1992intergenerational}. The resulting upward adjustment can be interpreted in two ways: either positively, as a correction for measurement error or transitory variation in income, or negatively, as a failure to account for the impact of idiosyncratic market shocks that are persistent within but not necessarily across generations, which effectively reduce the intergenerational transmission of income.

Second, mirroring the logic of the LATE theorem and our framework in Section \ref{sec:model}, a given IV estimator captures only the persistence linked to the variation in parental income induced by the chosen instrument(s). Since many individual characteristics that are commonly used as instruments, such as education, occupation, race, region, or combinations of these, tend to exhibit higher persistence across generations than income itself, TSLS estimates are typically biased upward relative to the true direct estimate.

However, this second factor can, in principle, also lead to a \textit{downward} bias relative to the direct parent-child IGE, such that IV estimates are not necessarily larger than OLS estimates. From eq. (\ref{beta_iv}), if the chosen instrument happens to be a determinant of income with comparatively low intergenerational persistence, then the resulting IV estimate will be \textit{smaller} than the direct estimate, as we illustrate below (and show formally in Appendix \ref{rel_lit}).\footnote{Moreover, \cite{bloise2021estimating} and \cite{cortes2025two} show that in the case of two-sample IV, the bias could also be negative due to out-of-sample prediction error (``overfitting'').} Our framework thus generalizes the earlier view that TSLS estimates provide an upper bound for the direct estimate, clarifying the conditions under which they overstate or understate the direct estimate. 

To illustrate our arguments empirically, we use Swedish administrative data that contain detailed information on labor incomes and other characteristics of parents and children. Appendix \ref{sec:swe_data} describes our sample selection and variables. We compare direct OLS estimates of the IGE and rank slopes using observed incomes with IV estimates using different parental characteristics as instruments. These IV estimates are estimated within the same sample, so we abstract from the additional issues that can arise in two-sample IV settings, as discussed in the next section.

Table \ref{iv_estimator} reports our results. Column (1) of the upper panel ("Men") shows that the direct OLS estimate is about 0.2, or slightly higher for the rank slope. Consistent with the prior literature, commonly used instruments such as education (col. 2) or occupation (col. 3) result in IV estimates that are higher than the direct estimates, though not dramatically so. However, using region of residence (col. 4) or country of birth (proxy for immigrant/ethnic background; col. 5) as instruments yields substantially higher estimates. In light of our theoretical arguments, this pattern is unsurprising, since region of residence and ethnicity are characteristics that tend to be more persistent across generations than education or the choice of occupation.
 
A related observation is the markedly lower first-stage $R^2$ seen in Columns (4) and (5). Naturally, instruments with low predictive power in the first stage can deviate more from the direct estimator. Conversely, as the $R^2$ approaches one, the (one-sample) IV estimate must eventually approach the weighted average underlying the direct estimate, as reported in Column (1).\footnote{A similar argument is made by \citealt{bjorklund1997intergenerational} and \citealt{jerrim2014two} in the case of two-sample IV, though the conditions for unbiasedness are more complex in that setting.} Similarly, maximizing the out-of-sample predictive power of the first stage can be an effective strategy to reduce bias in two-sample IV settings (\citealt{bloise2021estimating}). Still, the first-stage $R^2$ is only part of the story. With any imperfect first-stage prediction, the crucial question is whether the instrument captures income-generating characteristics that are themselves more or less persistent across generations compared to income itself, as given by the direct IGE. 

We can exemplify by comparing the two most common IV approaches in the prior literature, using either education (col. 2) or occupation (col. 3) as instruments. While both are highly imperfect predictors of income, with $R^2$ lower than 0.3, using education results in a \textit{higher} IV estimate. From the perspective of our model, the likely explanation is that education is more intergenerationally persistent than occupation. We cannot directly estimate the persistence of characteristics that enter non-parametrically (such as occupation dummies).\footnote{Moreover, some variables—most notably occupation codes—change in content and detail across generations.} As a workaround, we estimate persistence in the fitted values from separate earnings regressions on each characteristic. Appendix Table \ref{igps} shows that using this approach, the earnings advantages related to education seem indeed more persistent than those related to occupations. Consistent with Table \ref{iv_estimator}, we also find strong persistence in country of birth and, in particular, region of residence.\footnote{Note that this approach using fitted values is also sensitive to changes in the returns to characteristics across generations. For example, an estimate might be low not only because the intergenerational persistence of that characteristic is low, but also because the earnings returns to the characteristics changed between generations.}

As a further case in point, Column (8) illustrates what happens when we use as instrument the \textit{residual} variation in parental income that is not explained by any of the characteristics in Columns (2)-(5). Because this residual variation also reflects random determinants of earnings (such as market luck) and transitory fluctuations, we would expect it to be less intergenerationally persistent. In line with this argument, the IV estimates are now lower than the direct estimates in Column (1): the IGE drops from 0.199 to 0.156, and the rank slope from 0.231 to 0.191.
 
For a subsample of fathers we also observe cognitive and non-cognitive skill scores from military enlistment records.\footnote{Several prior studies use these data and provide further background, including \cite{lindqvist2011labor} and \cite{nybom2017distribution}. See Appendix \ref{sec:swe_data} for details.} Reestimating the same specifications for this subsample generally results in lower estimates, including for the direct estimates (Panel B, ``IQ sample'').\footnote{Note that the IV estimates using country of birth as instrument are unreliable and imprecise for this subsample, since very few immigrant fathers are enlisted in the military.} Employing these skill measures as instruments (Columns 6-7) yields estimates that are higher (cognitive) or substantially higher (noncognitive) than those based on education or occupation. Thus, we expect the intergenerational persistence of such skills (and their correlates) to be strong, and in particular stronger than for education or occupation. Indeed, this is what we find (Appendix Table \ref{igps}).\footnote{The implied strong persistence of non-cognitive skills may be surprising, but is broadly consistent with recent studies that adjust for measurement error in skills, such as \cite{gronqvist2017intergenerational}.}

\begin{table}[ht]
\centering
\begin{center}
\centering{}
{ 
\global\long\def\sym#1{\ifmmode^{#1}\else$^{#1}$\fi}%
\footnotesize
\begin{tabular}{lcccccccc}
\toprule
 & \multicolumn{1}{c}{OLS} & \multicolumn{1}{c}{IV} & \multicolumn{1}{c}{IV} & \multicolumn{1}{c}{IV} & IV & IV & IV & \multicolumn{1}{c}{IV}\tabularnewline
  & \multicolumn{1}{c}{} & \multicolumn{1}{c}{Education} & \multicolumn{1}{c}{Occupation} & \multicolumn{1}{c}{Region} & Birth country & Cognitive & Noncogn. & \multicolumn{1}{c}{Residual}\tabularnewline
 & (1) & (2) & (3) & (4) & (5) & (6) & (7) & (8)\tabularnewline
 \hline
 \addlinespace[1ex]
\multicolumn{9}{c}{\underline{Panel A: Men}}  \tabularnewline
 \addlinespace[1ex]
IGE & 0.199\sym{{*}{*}{*}} & 0.288\sym{{*}{*}{*}} & 0.255\sym{{*}{*}{*}} & 0.380\sym{{*}{*}{*}} & 0.383\sym{{*}{*}{*}} &  &  & 0.156\sym{{*}{*}{*}}\tabularnewline
 & (0.002)  & (0.004)  & (0.003)  & (0.012)  & (0.009)  &  &  & (0.002) \tabularnewline
$R^{2}_{\textit{first}}$ &  & 0.146 & 0.284 & 0.018 & 0.037 &  &  & 0.644\tabularnewline
 \addlinespace[1ex]
Rank slope & 0.231\sym{{*}{*}{*}} & 0.335\sym{{*}{*}{*}} & 0.289\sym{{*}{*}{*}} & 0.460\sym{{*}{*}{*}} & 0.381\sym{{*}{*}{*}} &  &  & 0.191\sym{{*}{*}{*}}\tabularnewline
 & (0.002)  & (0.004)  & (0.003)  & (0.011)  & (0.009)  &  &  & (0.002) \tabularnewline
$R^{2}_{\textit{first}}$ &  & 0.183 & 0.324 & 0.022 & 0.031 &  &  & 0.454\tabularnewline
$N$ & 395,424  & 395,424  & 395,424  & 395,424  & 395,424  &  &  & 395,424 \tabularnewline
 \addlinespace[1.5ex]
\multicolumn{9}{c}{\underline{Panel B: Men, IQ sample}} \tabularnewline
 \addlinespace[1ex]
IGE & 0.179\sym{{*}{*}{*}} & 0.257\sym{{*}{*}{*}} & 0.237\sym{{*}{*}{*}} & 0.329\sym{{*}{*}{*}} & 0.282\sym{{*}{*}{*}} & 0.326\sym{{*}{*}{*}} & 0.478\sym{{*}{*}{*}} & 0.138\sym{{*}{*}{*}}\tabularnewline
 & (0.003)  & (0.009)  & (0.006)  & (0.022)  & (0.049)  & (0.010)  & (0.014)  & (0.004) \tabularnewline
$R^{2}_{\textit{first}}$ &  & 0.122 & 0.226 & 0.019 & 0.004 & 0.088 & 0.055 & 0.697\tabularnewline 
 \addlinespace[1ex]
Rank slope & 0.209\sym{{*}{*}{*}} & 0.313\sym{{*}{*}{*}} & 0.264\sym{{*}{*}{*}} & 0.441\sym{{*}{*}{*}} & 0.292\sym{{*}{*}{*}} & 0.366\sym{{*}{*}{*}} & 0.495\sym{{*}{*}{*}} & 0.174\sym{{*}{*}{*}}\tabularnewline
 & (0.003)  & (0.008)  & (0.006)  & (0.021)  & (0.055)  & (0.009)  & (0.013)  & (0.004) \tabularnewline
$R^{2}_{\textit{first}}$ &  & 0.139 & 0.260 & 0.021 & 0.003 & 0.104 & 0.061 & 0.507\tabularnewline
$N$ & 104,405  & 104,405  & 104,405  & 104,405  & 104,405  & 104,405  & 104,405  & 104,405 \tabularnewline
\bottomrule
\end{tabular}}
\end{center}
\caption{Illustration of TSLS Estimates with Different Instruments}
\label{iv_estimator}
\vspace{-0.6cm}
\floatfoot{\footnotesize \textit{Note:} OLS and IV regressions in Swedish administrative registers.  Column (1) shows the ``direct'' OLS regression of son's on father's earnings. Columns (2)-(8) show IV estimates and the $R^2$ from the first stage using different sets of instruments for father's earnings: in col. (2) dummies for (father's) education level, in (3) 3-digit occupation dummies, in (4) region of residence dummies, in (5) dummies for country of birth, and in (6) and (7) measures of standardized cognitive and noncognitive skill from the military draft. In Panel B, all specifications are reestimated using the subsample with available draft data (``IQ sample''). In Column (8) we use as IV the residual from a joint regression of father's earnings on all instruments from Columns (2)-(7). All specifications are estimated both using log earnings (``IGE'') and earnings ranks (``Rank slope''). See Appendix \ref{sec:swe_data} for details.}
\end{table}
 
A framework incorporating ``multiplicity'' thus provides a useful lens for interpreting the TSLS approach in the intergenerational mobility context. It helps explain key findings from the existing literature and illustrates potential mechanisms underlying the observed biases. While TSLS estimators tend to diverge from direct estimates, they nonetheless capture meaningful and interpretable parameters that reflect specific aspects of the intergenerational transmission process.

\subsection{The Two-sample IV Approach}
\label{sec:TSLS}

While the previous subsection assumed a single sample containing both parent and child incomes, many data sources lack parental income data entirely. In such cases, researchers often use two-sample TSLS (TSTSLS) estimators, in which the first stage—the prediction of parental income from instruments such as education or occupation—is estimated in an auxiliary sample, and the predicted values are then linked to the main sample of parents and children through the shared instruments.
This approach is heavily relied upon in the empirical literature, with early examples including \citet{bjorklund1997intergenerational} and \citet{nicoletti2008intergenerational}.\footnote{The TSTSLS estimator was originally introduced by \citet{klevmarken1982missing}, as a tractable variant of the two-sample instrumental variables (TSIV) estimator (for discussions, see \citealt{arellano1992female} and \citealt{angrist1992effect}). TSTSLS and TSIV have the same probability limit, but can differ in small samples \citep{inoue2010two}. We focus here on TSTSLS, primarily since it has been the most common approach in the mobility literature.} Indeed, for most countries around the world, this is the only feasible way to estimate earnings or income mobility; accordingly, a large share of the country-specific estimates in prominent comparative surveys are based on a two-sample IV approach \citep{corak2013income,narayan2018fair,munoz2025intergenerational}. It is particularly relevant for developing countries and for historical analyses of
mobility trends \citep{bloise2021estimating}. 

However, the two-sample approach introduces additional methodological challenges that have sparked debate about its empirical validity. Concerns related to instrument excludability—or, in our framing, the role of instrument-specific and heterogeneous rates of persistence— apply equally to TSTSLS and may lead to upward-biased estimates. In addition, a concern specific to this approach is measurement error in predicted parental incomes, arising from the use of an external sample to estimate the first stage \citep{inoue2010two,bloise2021estimating}.\footnote{Prior work recognizes related issues. \citet{nicoletti2008intergenerational} proposed criteria for selecting first-stage predictors, and \citet{jerrim2014two} propose conditions under which TSTSLS is consistent (perfect first-stage prediction of parental income or no instrument endogeneity). However, they also note that increasing the first-stage predictive power (e.g., by incorporating more variables) could in principle worsen the bias if the added instruments are particularly endogenous. This argument mirrors the implications of our framework: adding a strongly intergenerationally persistent characteristic to the set of instruments could both raise the first-stage $R^2$ and the upward bias of the IGE. Along similar lines, \citet{cortes2025two} demonstrate that the TSTSLS estimator converges to the direct estimator when the explained variance in parental income matches the explained endogeneity in the child income equation.}  

These issues are reiterated by \citet{jacome2025mobility}, who emphasize three key challenges relevant for TSTSLS. First, recall bias can be a particular concern when children in the estimation sample provide information about their parents, such as their occupation or education, which they remember inaccurately. Second, the first-stage estimation is conducted in finite samples, introducing noise into the prediction. Third, differences in the data-generating processes between the first-stage and main samples might result in an incorrectly estimated conditional expectation.

We formalize these issues by writing the conditional expectation of parental income in the auxiliary sample as:
\begin{align}
     E_{2}[y_{t-1}\mid Z_{it-1}]
   =\sum_{j=0}^{J}\rho_{j}E_{2}[X_{t-1}^{j}\mid Z_{it-1}]
   =\sum_{j=0}^{J}\rho_{j}\left(E[X_{it-1}^{j}\mid Z_{it-1}]+\nu^{j}_{it-1}\right)
\end{align}
where $\nu^{j}_{it-1}
=E_{2}[X_{t-1}^{j}\mid Z_{it-1}]
- E[X_{it-1}^{j}\mid Z_{it-1}]$ captures the discrepancy between the two samples, arising from recall bias, finite-sample issues, or discrepancies in the conditional expectation between the two samples.\footnote{We define the conditional expectation in the second sample $E_{2}[X_{it-1}^{j}|Z_{it-1}]$ as the conditional expectation in the main sample $E[X_{it-1}^{j}|Z_{it-1}]$ plus an error term $\nu^{j}_{it-1}$.} Even under the assumption that the instrument perfectly matches the weighting structure of income (i.e. $\omega_{j}^{Z}=V(E[X^{j}_{it-1}|Z_{it-1}])=V(X^{j}_{it-1})=\omega_{j}$), the presence of first-stage errors from the use of the auxiliary sample introduces an additional term, $\omega^{Z}_{\nu_{j}}>0$, which attenuates the estimate relative to the direct IGE.

This attenuation bias can be expressed more explicitly in the second stage as follows:
\begin{align}
\label{bias_two}
\beta_{\textit{TSTSLS}}=\frac{Cov(y_{it};E_{2}[y_{t-1}|Z_{it-1}])}{V(E_{2}[y_{t-1}|Z_{it-1}])}=\frac{\sum_{j=0}^{J}\lambda_{j}\omega_{j}^{Z}}{\sum_{j=0}^{J}\left(\omega_{j}^{Z}+\omega^{Z}_{\nu_{j}}\right)}
\end{align}
where compared to eq. (\ref{beta_iv}) we now have $\omega^{Z}_{\nu_{j}}$ in the denominator, which biases the estimate towards zero. The magnitude of this attenuation bias is determined by $\omega^{Z}_{\nu_{j}}=\rho_{j}^{2}V(\nu^{Z}_{j})$ and thus depends on the variance of the error. 

Hence, the likelihood of attenuation bias increases in contexts where the child recalls the parental characteristics less accurately or where the first-stage estimation is noisier. For example, when using educational attainment as an instrument, the conditional expectation of income based on education is relatively straightforward, corresponding to the average income for individuals with a certain level of education. Since educational attainment is usually a low-dimensional variable, researchers require fewer observations to accurately approximate the conditional expectation, and differences in sampling between the estimation and main samples are unlikely to significantly affect the conditional expectation in this case. 

In contrast, when using fine-grained occupational categories as imputation variables, the situation could become more challenging. For rare occupations, the average income may be estimated with considerable noise in smaller samples, or the sampling differences between the two datasets might introduce greater misspecification in the conditional expectation. Such concerns may render the TSTSLS estimates more susceptible to attenuation bias. As discussed in Section \ref{sec:names}, this issue becomes particularly severe when using name-based instruments.

This leads to the question of how researchers should select their instruments in order to minimize attenuation bias, and better approximate direct estimates of income mobility. \cite{bloise2021estimating} provide a detailed comparison of different strategies, showing that machine learning algorithms can be used to minimize out-of-sample prediction error in the parental income imputation, or to choose across different candidate specifications.\footnote{Regularization and model selection in the first step may introduce bias in a second-step estimation, which could be addressed using the results from the literature on debiased machine learning \citep{chernozhukov2022locally,escancianoterschuur}.}

\subsection{Lessons from the Empirical Literature}
\label{subsec:literature_TSIV}
Despite these theoretical concerns, IV approaches offer practical solutions to data limitations and remain widely used in the mobility literature. In a classic application, \citet{solon1992intergenerational} showed that in a one-sample U.S. setting, TSLS produces estimates that are larger than the direct IGE. While Solon used parents' education as instrument for their earnings, \citet{zimmerman1992regression} relied on an index of socioeconomic status and found largely similar results. To what extent these elevated estimates are due to non-excludable instruments or a reduction in measurement error is unclear. Similarly, \citet{bjorklund1997intergenerational} observed that while intergenerational persistence in Sweden is lower than in the US, TSTSLS estimates for Sweden were higher than those derived using OLS. Comparable results have been reported for the UK by \citet{dearden1997intergenerational}.

\citet{jerrim2014two} document the extensive use of TSTSLS across various contexts. Before administrative data became more widely available for mobility studies, TSTSLS was commonly applied even in high-income countries. Early applications in the U.S. include
\citet{aaronson2008intergenerational}, who estimate intergenerational mobility over the second half of the 20th century by imputing parental income based on age and state of residence. In Europe, notable studies include those on France \citep{lefranc2005intergenerational,lefranc2010intergenerational}, the UK \citep{nicoletti2008intergenerational, bidisha2013microeconometric}, Spain \citep{cervini2015intergenerational}, Switzerland \citep{bauer2006intergenerational}, and Italy \citep{mocetti2007intergenerational, piraino2007comparable, barbieri2020intergenerational}. The method has also been applied in Canada \citep{fortin1998intergenerational}, Australia \citep{leigh2007intergenerational}, and Japan \citep{ueda2009intergenerational,lefranc2010intergenerational}. The TSTSLS-based estimates tend to be higher than direct estimates from linked administrative data that subsequently have been produced for some of these countries, again reflecting the concern about upward bias \citep{bjorklundjannti2009, chetty2014land, acciari2022and}. Despite the increased availability of administrative register data, the method remains highly relevant for countries where such registers are unavailable or do not allow direct linking of parents and children (e.g. for France, \citealt{kenedi2023intergenerational}).

More broadly, the TSTSLS method has become a cornerstone of large-scale cross-country comparisons of intergenerational mobility, particularly in data-scarce environments. Despite its well-known limitations, TSTSLS remains a valuable tool for generating suggestive evidence in contexts lacking comprehensive panel data, especially in developing countries. Prominent comparative studies led by international institutions, such as \citet{narayan2018fair} and \citet{munoz2025intergenerational}, rely heavily on this approach, as do many country-level analyses.\footnote{Examples from Asia include studies on China \citep{gong2012intergenerational}, Taiwan \citep{sun2015intergenerational}, and Korea \citep{ueda2013intergenerational}. In South America, researchers have applied TSTSLS to Brazil \citep{dunn2007intergenerational, ferreira2006intergenerational} and Chile \citep{nunez2010intergenerational} among others. For Africa, the method has been employed in South Africa \citep{piraino2015intergenerational} and more recently in Ghana \citep{opoku2024intergenerational}.} Across studies, a recurring finding is that intergenerational mobility tends to be lower in developing countries compared to developed ones, even after accounting for the upward bias inherent in TSTSLS estimates. This finding aligns with the observation that countries with higher inequality tend to exhibit lower mobility (dubbed the ``Great Gatsby Curve'', \citealt{hassler2007inequality,blanden2013cross,corak2013income,durlauf2022great}).

Imputation techniques play a critical role in quantifying mobility also beyond two-sample or regular TSLS, in contexts where income data—for either parents or children—are inadequate or missing.\footnote{Income data may also be incomplete if child income is observed at a young age or parental income at older ages, leading to poor approximations of lifetime income.} In historical contexts, the study of intergenerational mobility has been greatly advanced by the ability to link individuals across U.S. censuses.\footnote{The development of large-scale historical linkages \citep{feigenbaum2016automated,bailey2020automated,song2020long, abramitzky2021automated,Helgertz:2022,buckles2023breakthroughs} has sparked a growing literature on intergenerational mobility, data quality, and demographic dynamics.} In the absence of reliable income data, however, researchers typically focus on occupational mobility or impute income based on observable proxies such as occupation and race.\footnote{A notable exception is Iowa, where historical income records allow for direct measurement of intergenerational income mobility \citep{parman2011american, feigenbaum2018multiple}. A large body of work in both economics and sociology has analyzed long-run trends in occupational mobility using historical data \citep{guest1989intergenerational, ferrie2005history, long2013intergenerational, song2020long}.} For instance, \citet{abramitzky2021intergenerational} find that second-generation migrants historically exhibited greater mobility than natives, driven by favorable geographic selection. \citet{collins2022african} document persistent gaps in mobility between Black and White Americans throughout the 20th century. Correcting for measurement error and race-related sample selection bias, \citet{ward2023intergenerational} recovers higher historical persistence estimates than previously reported for the US.

A recent study employing a TSTSLS-like approach is \citet{jacome2025mobility}, which investigates mobility in the U.S. for cohorts born between 1910 and 1970, also including historically underrepresented groups such as African Americans and women in their analysis. The main finding is that mobility improved between the 1910 and 1940 cohorts, primarily due to a narrowing racial gap. Their methodology departs somewhat from standard TSTSLS, as they first impute parental income using occupation, race, and geography before applying a logarithmic transformation. Nonetheless, they assess the robustness of their results to the typical concerns associated with TSTSLS, such as attenuation bias from imperfect first-stage predictions and concerns about endogeneity. 

These contributions illustrate a broader and rapidly expanding research agenda on historical mobility, largely enabled by the use of imputation and instrumental-variable strategies to overcome severe data limitations. However, as these methods continue to push the global and temporal scope of mobility research, careful analysis of their identifying assumptions and potential biases becomes increasingly important.

\section{Name-Based Estimators}\label{sec:names}

A more recent strand of research uses \textit{names} to study intergenerational transmission. Motivated by their role as probabilistic markers of family ties, names are now used in a variety of ways, both in methods and research objectives, across studies and disciplines.

The literature can be broadly divided into two main approaches. The first approach exploits surnames in a cross-sectional setting, without explicitly linking parents and children. The key idea is that the extent to which surnames predict socioeconomic outcomes within a generation—what \cite{guell2015informational} define the \emph{Informational Content of Surnames} (ICS)—is informative about intergenerational persistence. In a society with perfect mobility, surnames would carry no predictive power, whereas in a fully immobile society they would be highly informative.\footnote{This predictive power of surnames may itself reinforce intergenerational inequality if decision-makers or AI systems use them as proxies for skills (\citealt{algorithmicnames}).} Hence, the $R^{2}$ from a regression of incomes on surname indicators can be linked to the intergenerational income elasticity (IGE) or correlation. Building on this idea, \cite{guell2018correlating} study variation in persistence across Italian provinces and relate it to local economic conditions.

The second approach uses names as instruments—both first names (\citealt{olivetti2015name}) and, more commonly, surnames.\footnote{\cite{olivetti2015name} estimate intergenerational elasticities between fathers and children of both sexes in the U.S. by exploiting the socioeconomic information embedded in first names to construct pseudo-links across generations. \citet{olivetti2018three} extend this approach to study mobility over three generations, while \citet{olivetti2024emergence} apply it to analyze the role of family status in the U.S. marriage market.} This approach is motivated by two main reasons. First, names can be used to link individuals probabilistically across generations when direct links are unavailable (\citealt{collado2012long}), or to study much more distant generations than standard data allow. For example, \citet{barone2021intergenerational} use surnames to link individuals in 15th-century Florence to their modern-day pseudo-descendants, showing that socioeconomic advantages persist over more than five centuries. More recently, studies such as \citet{ager2021intergenerational} and \citet{althoff2024jim} use surname-based approaches to examine how historical events continue to shape socioeconomic outcomes in more recent times.\footnote{Several studies use names to examine intergenerational transmission of status across a range of contexts. In historical settings, \citet{belloc2024multigenerational} analyze wealth transmission in Renaissance Florence, while \citet{haner2024name} investigate social mobility in Switzerland. In developing countries, names often serve as a proxy in the absence of detailed data, for example, in Colombia \citep{alvarez2023persistence, jaramillo2025does}, in both Colombia and Chile \citep{jaramillo2021surnames}, in Korea \citep{paik2014does}, in Hungary \citep{bukowski2022social} and in Sierra Leone \citep{dupraz2024elite}.}  

Second, surnames are deliberately used to correct for measurement error, beyond simply overcoming data limitations. \citet{clark2014also} and \citet{clark2015intergenerational} propose a latent factor model in which transmission occurs through an underlying, unobserved status component, corresponding to the ``simple'' latent factor model discussed in Section \ref{sec:model}. If income or other observed status measures are noisy proxies for this latent status, standard estimates suffer from attenuation bias. Studying the persistence of average status at the surname level would reduce this bias, effectively using surnames as instruments in a TSLS approach (\citealt{SantarvitaStuhlerEJ}). Through this approach, Clark and co-authors find very high persistence rates, often in the range of 0.7–0.8, well above typical parent-child estimates.\footnote{From this, they formulate a provocative ``Law of Social Mobility'', suggesting that socioeconomic persistence is both high and stable across societies and time.} 

However, long-run mobility studies and surname-based analyses may themselves suffer from bias. First, name-based estimators may suffer from weak-instrument problems and are highly sensitive to sampling properties \citep{SantarvitaStuhlerEJ}; we return to these issues in Section \ref{subsec:names_TSIV}. Second, as emphasized by \cite{croix2024nepotism}, bias may also stem from sample selection. Many applications rely on samples that are not fully representative of the underlying population—for example, when focusing on specific regions and excluding migrants (e.g., \citealt{barone2021intergenerational}), or when using data sources that only include individuals meeting certain conditions, such as probate or wills records (\citealt{clark2015intergenerational}). While we do not address these issues in detail, they may be highly relevant in applications.

While name-based studies have been particularly popular in economics, their use extends across the social sciences. For instance, \cite{torche2018estimating} distinguish between individual, within-group, and between-group mobility, and argue that the surname-based approach used by Clark and co-authors confounds individual-level transmission with between-family heterogeneity. Similarly, \cite{dalman2025legacy} combines surname information with observed parent-child links to separate within-family from surname-group persistence. Using Swedish data, she finds that surnames themselves reflect a heritable group-level status dimension, which persists strongly across generations. 

Name-based studies thus provide novel insights, but have also sparked considerable debate. A central question is whether surname-based designs serve as substitutes for direct intergenerational measures or instead capture fundamentally different parameters. In this section, we use the multiplicity framework to better understand what surname-based estimators capture. Drawing on \cite{delpizzo2025in}, we argue that surname-based designs identify a different parameter than direct measures such as the IGE. While they mitigate attenuation bias, they also place greater weight on highly persistent transmission pathways than conventional parent-child correlations. Because these pathways account for the bulk of multigenerational correlations, surname-based estimators are especially informative about \textit{long-run} dynamics, but less suited to capturing \textit{short-run} (parent-child) persistence. Our discussion here complements a recent review by \cite{SantarvitaStuhlerEJ}, which focuses on the influence of weak instruments and sampling properties on name estimators (see Section \ref{subsec:names_TSIV}).

\subsection{The Interpretation of Name-Based Estimators}\label{subsec:name_interpretation}

We begin by clarifying the parameter identified by surname-based estimators: are they substitutes for conventional parent-child estimates, or do they identify a fundamentally different parameter? 

To address this question, we analyze their theoretical properties within our framework using a simple two-stage least squares (TSLS) specification. The use of non-overlapping samples in a two-sample IV approach generates additional complications, which we review in Section \ref{subsec:names_TSIV}. In the surname context, equation \eqref{beta_iv} can be recast as follows:
\begin{equation}\label{beta_names}
    \beta_{\textit{TSLS}}=\frac{Cov(y_{it};E[y_{it-1}|S_{it-1}])}{V(E[y_{it-1}|S_{it-1}])}=\frac{\sum_{j=0}^{J}\lambda_{j}\omega_{j}^{S}}{\sum_{j=0}^{J}\omega_{j}^{S}+\omega_{u}^{S}}
\end{equation}
where the weighting structure can be expressed as:
\begin{align}
    \omega^{S}_{j}=\rho_{j}^{2}V(E[X_{it-1}^{j}|S_{it-1}])+\rho_{j}\sum_{k\neq j}^{J}\rho_{k}Cov(E[X_{it-1}^{j}|S_{it-1}],E[X_{it-1}^{k}|S_{it-1}])
\end{align}
Relative to parent-child correlations, surname-based estimators differ along two key dimensions. First, surname-based estimators average out ``measurement error'' more than standard parent-child estimates.\footnote{Here, “measurement error” is used broadly to include both misreporting of observed outcomes and the fact that even perfectly measured variables may be imperfect proxies for an underlying “generalized social status,” as emphasized by Clark and co-authors.} This argument can be illustrated using our framework in Section \ref{ind_estimator}, with the characteristics $X^j$ representing a person's latent status and $u$ the measurement error. Averaging within surnames reduces both $\omega_j^S$ and $\omega_u^S$, but when the error is classical and independent across individuals, the reduction is stronger for the error component, since it does not share a common component within surnames. As a result, attenuation bias is reduced, and the contribution of the error term in eq.~\eqref{beta_names} is smaller than that of the corresponding individual-level term in eq.~\eqref{eq:corr2g}.\footnote{This is analogous to the ``conventional'' TSLS approach with education or occupation in Section \ref{sec:IV}, where measurement error is similarly reduced by averaging over education or occupation groups.}

However, reduced measurement error is not the only difference between surname-based and conventional parent-child correlations. A second difference, related to our discussion of traditional IV approaches above, lies in how these estimators weight underlying traits and transmission pathways. As in standard TSLS, surname-based estimators place greater weight on characteristics that are well predicted by the instrument. In this context, traits that are well predicted by surnames are precisely those that are shared within surname groups and differ across them. We develop this argument in more detail in \cite{delpizzo2025in}, but provide a summary here.

To illustrate, consider a trait such as  ``grit''. If individuals with high and low grit are similarly distributed across surnames, then knowing someone’s surname tells us little about that trait, and the between-surname variance in grit is very small. As a result, surname-based estimators would assign little weight to this trait. By contrast, traits that are strongly clustered within surnames, such as geographical location, receive much greater weight, even if those traits play only a limited role in intergenerational transmission. In the extreme, if all individuals sharing a surname live in the same location, knowing the surname is equivalent to knowing their location: the surname becomes a perfect predictor of geography.

Existing explanations for the gap between surname-based and direct estimates map naturally into these two channels: reduced attenuation and a different weighting structure. \citet{clark2014also} emphasizes the first channel. He attributes this gap to measurement error in a broad sense, arguing that observed outcomes are imperfect proxies for an underlying ``generalized'' social status, which surname-based methods capture more accurately.\footnote{In contrast, \citet{vosters2017intergenerational} show that aggregating standard \textit{observable} socioeconomic indicators does not, by itself, reproduce the high persistence estimates reported by Clark and co-authors. This suggests that the large estimates reported by Clark and co-authors cannot be explained solely by averaging over noisy observables, but instead require that surnames capture more persistent underlying components of status. This divergence may arise if important determinants of outcomes are inherently unobserved (“latent”) and are better captured by the grouping approach.} By contrast, other authors argue that the high persistence captured by surname-based estimators reflects group-level differences rather than individual-level transmission (\citealt{torche2018estimating}). In our framework, this corresponds to the second channel: they place greater weight on group-level characteristics than parent-child correlations do. For instance, \citet{chetty2014land} argue that in the U.S. this pattern reflects the correlation between surnames and race, while \citet{guell2018correlating} show that in Italy it reflects persistent regional differences.
 
But what determines these weights, and why do they differ from those underlying direct parent-child correlations? The key insight is that surnames capture not just immediate family ties (e.g., father-child), but also more distant kinship links, such as cousins or uncles. We formalize this in a stylized single-parent setting, assuming that each surname group descends from one common ancestor whose characteristics are transmitted to later generations with the same name. Thus, the correlation among surname-group members is rooted in the traits inherited from this distant ancestor.
Formally, we can express the between-surname variability of a given characteristic $X_{it-1}^{j}$ as follows (see Appendix \ref{proof_name}):
\begin{equation}\label{eq:varX_name_dist}
    V(E[X_{it-1}^{j}|S_{it-1}])=E[\lambda_{j}^{2(t-1-\tau_{s})}]V(X_{i\tau_{s}}^{j}).
\end{equation}
This expression can be decomposed into two components: the persistence rate for that characteristic, $\lambda_{j}$, and the distance to the common ancestor ($t-1-\tau_{s}$).\footnote{For simplicity, we abstract here from the covariances between different traits, although the reasoning can be easily extended to include them (see Appendix \ref{proof_name}). Moreover, our presentation here hinges on the assumption that the characteristics of the common ancestor $X_{i\tau_{s}}^{j}$ are independent of the generation in which the common ancestor lived ($\tau_{s}$). For a more thorough discussion, see \cite{delpizzo2025in}.} First, traits with higher parent-child persistence receive more weight in surname-based compared to direct estimates, as long as the father is not the common ancestor himself (i.e., the exponent is greater than zero).\footnote{This does not imply that $\lambda_{j}$ is irrelevant when the father is the common ancestor; rather, in that case, the surname-based estimate reflects the standard parent-child correlation, since the weights $\omega^{Sur}_{j}$ in eq. (\ref{beta_names}) are no longer amplified by generational distance.} For example, if a personality trait like grit is not strongly transmitted from parents to children (small $\lambda_{j}$), surnames will offer little predictive power. In contrast, for highly persistent traits like geography (high $\lambda_{j}$), surnames will remain informative.

Second, persistence compounds with genealogical distance. The further back the common ancestor, the more weight is placed on highly persistent traits. Low-persistence traits, like grit, may even be transmitted from parent to child, but fade quickly across generations. For instance, if $\lambda_j = 1/5$, the correlation in $X^j$ falls below 0.01 after only three generations. In contrast, for highly persistent traits like geography, even distant ancestors may still shape your location. For example, if $\lambda_j=4/5$, the correlation in $X^j$ falls below 0.01 only after twenty generations. 

Surname-based estimates thus tend to exceed direct estimates because they place greater weight on characteristics that persist strongly across generations.\footnote{Although surname-based estimates overweight high-persistence traits, they are not always higher than direct estimates in practice, as they may also suffer from weak-instrument bias. We return to this point in Section \ref{subsec:names_TSIV}.} This is similar in spirit to standard imputation methods, where instruments capture more persistent characteristics: in that setting, commonly used imputing variables such as education are more persistent than income; here, instead, the pattern follows directly from the structure of surname-based estimators.

We provide evidence on these patterns in the next two subsections: Section \ref{subsec:namefreq} documents how surname-based estimates exceed parent–child correlations and vary with surname size, while Section \ref{subsec:name_pathways} examines the underlying weighting structure and identifies the characteristics driving these patterns.

\subsection{Name-Based Estimates Increase with Surname Frequency}\label{subsec:namefreq}

Existing evidence on surname-based estimates is mixed. While the findings of Clark and co-authors suggest very high levels of intergenerational persistence, other studies find more modest estimates. One potential explanation for these differences is attenuation bias arising in settings with non-overlapping samples, as discussed in Section \ref{subsec:names_TSIV}. Importantly, this variation arises not only across contexts, but also within them—across surname groups. A consistent pattern across studies is that surname-based estimates tend to be lower for rare surnames and increase systematically with surname frequency (\citealt{chetty2014land,guell2018correlating}). 

This pattern is consistent with the two mechanisms in our model. First, averaging over more individuals within a surname group reduces idiosyncratic noise that is not transmitted across generations. Because this noise averages out faster than transmitted traits, attenuation bias declines (\citealt{clark2014also}). Second, under positive population growth, larger surname groups may reflect more distant ancestry. As genealogical distance increases, less persistent traits decay quickly, leaving only highly persistent traits shared among group members. By contrast, smaller surname groups may capture closer kinship, allowing even less persistent traits to remain correlated.  

To probe these predictions, we focus on a favorable data setting with ``overlapping'' samples, where attenuation bias is limited (see \citealt{delpizzo2025in} for detailed evidence). Specifically, we use data from males aged 30 to 40 in the 1940 U.S. Census linked to their fathers in the 1920 Census, and estimate the surname-group (or ``grouping'') estimator across samples that include progressively larger surname groups. We divide relatively rare surnames (fewer than 1,500 male working-age members) into deciles by group size. Starting with the smallest decile (groups of up to two members), we compute both estimators and progressively expand the sample by adding subsequent deciles.
\begin{figure}[t]
    \centering
    \includegraphics[width=0.85\linewidth]{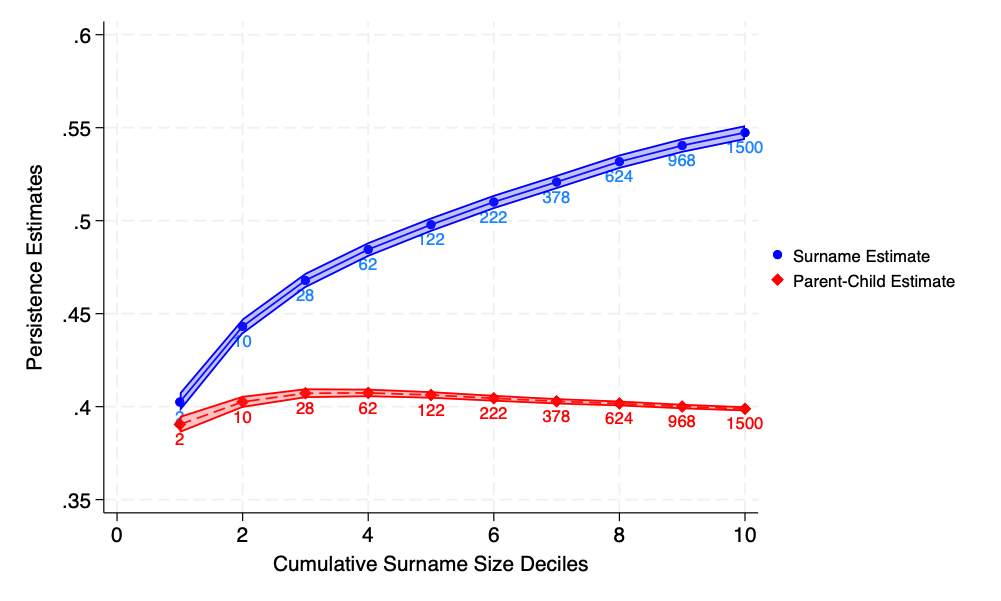}
    \caption{Grouping Estimator in U.S. Census Data (Occupational Score)}
    \label{surIGE}
\floatfoot{\footnotesize \textit{Note:} Surname-based grouping estimates (blue) and direct father-child estimates (red), computed on the same sample. The x-axis reports deciles of the surname size distribution (in 1920), restricting attention to relatively rare surnames. Each point includes surnames with size weakly below the cutoff corresponding to that decile; the numerical labels next to each point report these cutoffs (i.e., the maximum number of observations used to compute the surname average). The sample consists of male workers aged 30–40 in the 1940 U.S. Census linked to their fathers. The dependent variable is the log-occupational score. The regressor in the surname-based estimator is the surname-level average log-occupational score in 1920, computed over working-age males.}
\end{figure}

Figure \ref{surIGE} plots these grouping estimates and compares them to a direct measure of persistence, the father-child estimate in occupational scores, estimated in the same samples. The results support the predictions of the model. First, in the absence of weak-instrument bias from non-overlapping samples (see Section \ref{subsec:names_TSIV}), surname-based estimates exceed parent–child estimates (\citealt{clark2014also}). Second, grouping estimates increase with surname frequency (solid line), consistent with \citet{chetty2014land} and \citet{guell2018correlating}. Crucially, this gradient does not reflect differences in family-level transmission: the parent–child estimate (dashed line) remains essentially flat across the surname distribution.\footnote{Family characteristics do differ systematically with surname size, as described in Table \ref{tab:surname_freq_summary}, especially with a more urban population in rarer surnames. However, these differences are not large enough to have a meaningful influence on the parent-child estimate.}

We can thus conclude that, in the absence of substantial measurement error, surname-based estimators exceed parent–child correlations and increase with surname group size. However, these patterns do not by themselves reveal the underlying mechanisms. In particular, what is the weighting structure of surname-based estimators, and which characteristics drive these patterns? We address these questions in the next subsection.

\subsection{Which Mechanisms Do Surname Estimators Capture?}\label{subsec:name_pathways} 
What surname-based estimators capture is subject to debate. \citet{clark2014also} argues that surnames proxy for a generalized notion of social status, while others emphasize that they reflect group-level characteristics (\citealt{torche2018estimating, dalman2025legacy}), such as race (\citealt{chetty2014land}) and geography (\citealt{guell2018correlating}). Our framework offers a unifying perspective: surnames are more predictive of characteristics that are highly persistent across generations, and this is reflected in the estimator's implicit weights. Hence, examining how well surnames predict different observable characteristics can help identify which traits they capture in practice, and whether these align with interpretations proposed in the literature.

Moreover, we can exploit the gradient across surname group size to study the persistence of different characteristics. Because low-persistence traits are less well predicted in larger surname groups, smaller groups may place relatively more weight on such traits. Variation in these weights with surname size is therefore informative: characteristics whose weight increases with group size are more persistent, while those whose weight declines are less persistent. This approach is formalized in \cite{delpizzo2025in}. 

For illustration, we measure how strongly surnames predict different observable determinants of the outcome across surname size in the U.S. Census data.\footnote{In the model, the weights are defined using surname-level averages of parental characteristics in 1920. Here, instead, we use outcomes for the child generation in 1940 to construct analogous objects. The exercise is therefore meant to illustrate the patterns in the data, rather than to recover the model-based weights.} We first regress occupational scores (median income by occupation) on each characteristic separately and obtain fitted values for individuals in the 1940 child generation.\footnote{We also compute residuals from a joint regression of occupational score on state of residence, birthplace, education, urban/rural status, and race.} We then average both fitted values and outcomes at the surname level. Next, we group surnames into size deciles (restricting to surnames with fewer than 1,500 members) and progressively expand the sample to include larger bins. At each step, we compute the variance of surname-level averages of both fitted values and the outcome across included surnames. We approximate the weight of each characteristic as the ratio of these variances, normalized relative to the smallest size bin. Assuming each characteristic's relevance ($\rho_j$) does not vary with surname size, differences in these normalized ratios reflect differences in persistence ($\lambda_j$).

\begin{figure}[t]
    \centering
    \includegraphics[width=0.75\linewidth]{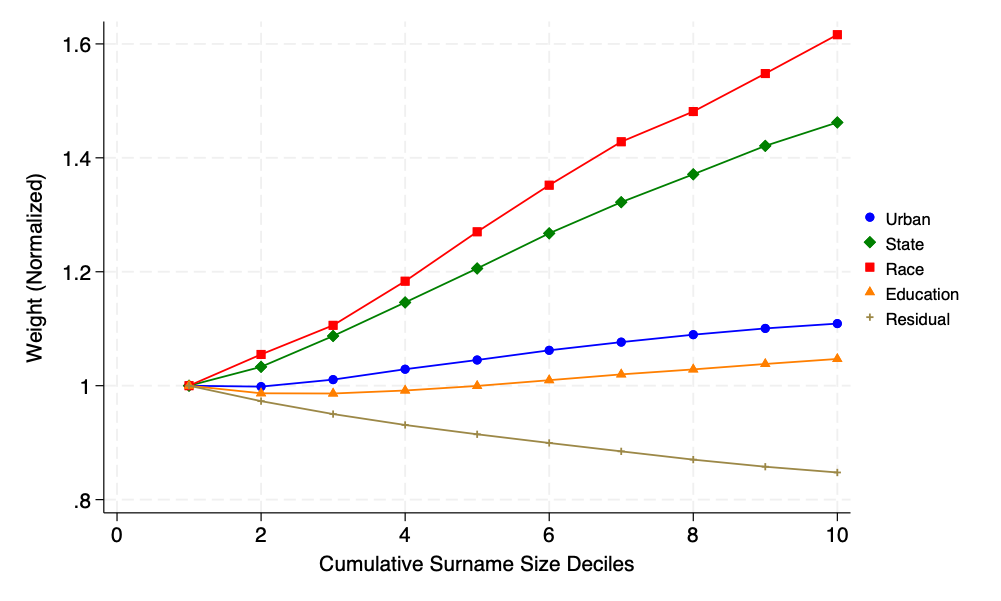}
    \caption{Weights of Different Characteristics in the Grouping Estimator}
    \label{Weights}
      \floatfoot{\footnotesize \textit{Note:} The figure shows how the empirical weight assigned to each characteristic varies across surname-size deciles (based on the 1920 distribution), focusing on relatively rare surnames. Each point corresponds to surnames with size weakly below the cutoff for a given decile. Weights are defined as the ratio of the variance of surname-level averages of fitted values (for a given characteristic) to the variance of surname-level averages of the overall outcome, and are normalized relative to the first decile. The sample consists of male workers aged 30–40 in the 1940 U.S. Census, linked to their fathers (\citealt{buckles2023breakthroughs}). The residual is obtained from a regression of occupational score on state of residence, birthplace, education, urban/rural status, and race.}
\end{figure}
As shown in Figure \ref{Weights}, we find that geographic location and race are among the most persistent observable characteristics, as their contribution to surname-level differences increases with surname size.\footnote{Our estimates regarding race should be interpreted with caution, due to the relatively small share of African-Americans in our sample. However, our results here align with the observation of enduring racial disparities in the U.S., and the finding that those disparities are a key contributor to high intergenerational persistence \citep{collins2022african,jacome2025mobility}.} These results support the view that name-based estimates reflect persistent ethnic gaps (\citealt{chetty2014land, dalman2025legacy}) and geographic differences (\citealt{guell2018correlating}). We also find that education and urban–rural status are more persistent than occupational scores, though less so than geography or race. By contrast, the residual component of occupational scores—including measurement error—is much less persistent than the outcome itself. Interestingly, these findings on the relative rates of persistence across traits are reminiscent of what we found in a completely different data set—based on modern Swedish administrative registers—in Section 3, suggesting that the patterns may be generalizable across time and space.

More broadly, our findings are consistent with both leading interpretations of high name-based estimates: surname-based estimators capture persistent environmental influences \citep{torche2018estimating, chetty2014land, guell2018correlating}, and reduce measurement error \citep{clark2014also}. Our framework, however, clarifies why this is the case: surnames place greater weight on traits that are themselves highly persistent. For example, surnames capture racial identity in U.S. data, and recent evidence shows that racial inequality is a key driver of long-run persistence (\citealt{jacome2025mobility, ward2025like}). The same logic helps explain why surname-based estimates tend to decline when conditioning on location, a comparatively persistent trait (\citealt{collado2012long, feigenbaum2018multiple, SantarvitaStuhlerEJ} or \citealt{haner2024name}).

\subsection{Do Surname Estimators Capture Long-Run Persistence?}\label{subsec:long_run}

Surname-based estimators thus ``overweight'' more persistent transmission pathways, but whether this constitutes a limitation depends on the research objective. If the goal is to understand long-run persistence, emphasizing these pathways may be a feature, not a flaw.

But what \textit{specific} parameter do surname-based estimators capture? Our results from Sections \ref{sec:model} and \ref{subsec:name_interpretation} imply that they identify the \emph{rate of decay}, $\beta_g / \beta_{g-1}$, at large genealogical distances $g$. The intuition is simple: surname-level averages reflect correlations between distant kin, which—in models with multiplicity—are mainly driven by the most persistent traits. Consequently, name-based estimators capture how these correlations decline across generations at long horizons.\footnote{More precisely, because the distance to a common ancestor varies, surname estimators capture a weighted average of decay rates across various generational distances.} In our framework, this object can be interpreted as a reduced-form measure reflecting a composite of the underlying factor-specific persistence rates $\lambda_j$.

This object is informative in its own right, as it provides a direct measure of long-run dynamics. In a simple latent-factor model, where observed outcomes are imperfect proxies for an underlying persistent trait, short-run persistence does not capture the true latent persistence parameter $\lambda$, whereas the decay rate at longer horizons may provide a closer approximation. In a multi-factor model with varying rates of persistence, the interpretation is not as neat; rather than ``true'' persistence, surname-based estimates capture a weighted average of decay rates across various generational distances. Because individuals in larger surname groups are unlikely to be closely related, these estimates primarily capture decay rates at higher genealogical distances, where we expect less rapid decay than at lower genealogical distances. 

These arguments imply that researchers can gain a more comprehensive understanding of mobility dynamics by combining different estimators. In particular, when long-run persistence $\beta_g$ cannot be directly observed—because data do not span many generations—it can be inferred by combining short- and long-run information. As discussed in Section \ref{sec:model}, $\beta_g$ reflects the accumulation of successive decay rates, so evidence from both margins helps pin down long-run persistence. Parent-child correlations capture short-run decay, while surname-based estimators capture decay at longer distances; together, they trace the evolution of persistence across generations.

The preceding discussion has focused on favorable data scenarios, in which the surname averages are constructed from large samples that cover the same families across generations. In the next section, we examine how name-based estimators behave in less favorable settings.

\subsection{Attenuation in Surname-based Estimators}\label{subsec:names_TSIV}

Our discussion so far explains why surname-based estimates can exceed direct parent-child estimates. Yet in some studies they are similar, or even smaller. Although this appears to contradict our interpretation, such variation largely reflects differences in empirical context: surname-based estimators are vulnerable to weak-instrument bias and highly sensitive to how surname averages are constructed (\citealt{SantarvitaStuhlerEJ}).

Two distinct mechanisms generate attenuation in surname-based estimates, and in practice they often operate jointly. The first reflects sampling noise: when surname averages are computed from finite samples rather than the full population, they become noisy measures of the underlying conditional expectation. The second reflects ``lack of overlap'' (\citealt{SantarvitaStuhlerEJ}), that is, when surname averages are constructed from samples that exclude individuals’ own parents. While sampling noise is always present in sampled data, lack of overlap can arise to different degrees depending on the empirical setting.

When only sampling noise is present—that is, with full overlap—attenuation pushes estimates toward the parent-child correlation. When lack of overlap is also present, attenuation is stronger, pushing estimates further down.

For clarity of exposition, we begin by isolating the first mechanism, although in practice it rarely arises without the second. When surname averages are computed from finite samples, they are noisy estimates of the conditional expectation $E[X_{it-1}^{j} \mid S_{it-1}]$. Formally, let $\overline{X}^{j}_{it-1}$ denote the estimated surname-level average. When calculated from a subsample, its variance combines the true between-surname variance and an additional sampling-error component:
\begin{equation*}
V(\overline{X}^{j}_{it-1})
= V\left(E[X_{it-1}^{j} \mid S_{it-1}]\right)
+ V(\nu^{j}_{it-1}),
\end{equation*}
where $V(\nu^{j}_{it-1})$ captures the sampling noise arising from small surname groups. As the number of observations per surname decreases, the sample mean becomes a noisier measure of the true population mean and increasingly reflects the father’s own outcome.

This implies that sampling noise generates attenuation toward the direct parent-child correlation. To see the intuition, consider the limiting case of a single observation per surname: the surname average coincides with the father’s outcome, and the estimator collapses exactly to the parent-child correlation. Consistent with this intuition, \citet{ward2023intergenerational} shows that estimates based on a 1\% Census linked sample increase substantially when instead the full census is used.

We provide empirical evidence for these mechanisms by comparing direct parent-child mobility estimates to surname-based estimates constructed from subsamples of the data. Specifically, we draw a random five-percent sample from our baseline dataset of males aged 30–40 in 1940 who are linked to their fathers in 1920 (with sampling performed at the level of linked father–son pairs). Within this sample, we construct surname averages using only the fathers observed in the sample. That is, for each surname, we average paternal outcomes across the subset of fathers included in the 5\% sample.\footnote{Because surname averages are computed using the same fathers that appear in the estimation sample, this setup implies full overlap between individual outcomes and surname-level averages.} For each individual, we therefore observe three measures: the father’s own outcome, the surname average computed within the same 5\% sample, and the surname average computed from the full census (the same measure used in Figure~\ref{surIGE}).

Table \ref{sampling} reports the results: Column (1) presents the direct parent-child estimate; Column (2) uses surname averages computed \textit{within} the sample; and Column (3) uses surname averages computed across the full census. As expected, the estimates in Columns (1) and (2) are very similar, reflecting the fact that with ``full overlap'' the surname-based estimator approaches parent-child estimates in small samples. In contrast, the estimate in Column (3) is substantially larger: because surname averages are computed from the full population, they are less affected by sampling noise and therefore provide a cleaner measure of between-surname variation. This pattern is consistent with the idea that surname averages constructed from the full population place greater weight on more persistent characteristics.

A second issue—'lack of overlap' (\citealt{SantarvitaStuhlerEJ})—arises when surname averages are constructed from a sample that excludes an individual’s own parent. This is common when parent and child samples are drawn independently, such as from separate Census extracts. In this case, the child’s outcome is no longer linked to an average over both close and distant relatives, but only to more distant ones, as the parent’s outcome is no longer included in the surname-level average. As a result, attenuation does not simply push the estimate toward the parent-child correlation, but further downward, toward the correlation between the child and more distant kin. To gain intuition, consider the extreme case in which there is only one observation per surname and it does not correspond to the child’s parent. For example, if the only individual sharing the surname is an uncle, the surname-based estimator effectively collapses to the correlation between the child and that relative.

We demonstrate the implications of this mechanism empirically. Starting from the 5\% sample of children observed in 1940 and linked to their fathers, we construct surname-based estimators using surname averages computed from an independent sample. Specifically, we draw a random five-percent sample of working-age males in 1920 and compute surname averages based on this sample. We then merge these surname averages to the 1940 child sample through the surname. Because the two samples are drawn independently, the fathers of the children in the estimation sample are unlikely to be included in the construction of the surname averages, implying very limited overlap. As shown in Column (4) of Table \ref{sampling}, this results in an estimate that falls well below the parent-child estimate in Column (1).
\begin{table}[H]
    \centering
\resizebox{0.85\textwidth}{!}{
{
\def\sym#1{\ifmmode^{#1}\else\(^{#1}\)\fi}
\begin{tabular}{lccccc}
\hline\hline 
            &\multicolumn{5}{c}{Dependent variable: Child Occscore}\\
            &\multicolumn{1}{c}{Direct} & & \multicolumn{3}{c}{Name-based} \\
            \cmidrule{2-2} \cmidrule{4-6}
            &\multicolumn{1}{c}{(1)}& &\multicolumn{1}{c}{(2)}&\multicolumn{1}{c}{(3)}&\multicolumn{1}{c}{(4)}\\
\hline
Father Occscore &       0.393\sym{***}&   &                  &                     &                     \\
            &    (0.00215)     &    &                     &                     &                     \\
[0.5em]
Surname Occscore &          &           &       0.444\sym{***}&                     &                     \\
\textit{in 5\% linked sample (full overlap)}             &          &           &    (0.00394)         &                     &                     \\
[0.5em]
Surname Occscore &           &          &                     &       0.593\sym{***}&                     \\
\textit{in full census}    &           &          &                     &     (0.00715)         &                     \\
[0.5em]
Surname Occscore&              &       &                     &                     &       0.286\sym{***}\\
\textit{in different 5\% sample (limited overlap)}          &         &            &                     &                     &     (0.00740)         \\
[0.5em]
\(N\)       &      223,729       &  &      223,729         &      223,729         &      196,915         \\
\hline\hline
\multicolumn{6}{l}{\footnotesize Standard errors in parentheses, \sym{*} \(p<0.05\), \sym{**} \(p<0.01\), \sym{***} \(p<0.001\)}\\
\end{tabular}
}
}
    \caption{Direct and Surname-based Grouping Estimators in a 5\% Sample of U.S. Census}
    \label{sampling}
  \floatfoot{\footnotesize \textit{Note:} The table compares direct and surname-based estimates of intergenerational persistence using a 5\% random sample of males aged 30-40 in the 1940 Census linked to their fathers. Column (1) reports the coefficient from a regression of the son’s log-occupational score on the father’s log-occupational score. Columns (2)-(4) report analogous regressions using surname-level averages as regressors. In Column (2), surname averages are computed across fathers within the same linked 5\% sample, implying full overlap between individual outcomes and surname-level averages. In Column (3), surname averages are computed across all working-age males from the full 1920 Census. In Column (4), surname averages are computed across fathers from an independent 5\% sample of working-age males in 1920, implying limited overlap. The smaller sample size in Column (4) reflects surnames in the child sample that are not observed in the independent 5\% 1920 sample.}
\end{table}

These results show that surname-based estimators are highly sensitive to the sampling design and, more generally, to how surname averages are constructed. Sampling noise attenuates estimates toward the direct parent-child estimate, while lack of overlap pushes them further down. This distinction is crucial for interpreting the empirical evidence: it helps explain why surname-based estimates vary widely across studies, depending on data quality, sample size, and the degree of overlap between samples.

\subsection{Regional Intergenerational Mobility with Surnames}

Even if name-based approaches identify a different estimand than traditional parent-child correlations, they may still serve as substitutes for standard measures for \textit{comparative} analyses. For instance, a large literature examines regional differences in intergenerational mobility (\citealt{chetty2014land, chetty2018impacts})\footnote{Examples include \citet{heidrich2017intergenerational}, who documents regional variation in income mobility within Sweden; \citet{fletcher2019intergenerational}, who examine geographic and temporal patterns in U.S. educational mobility; \citet{connolly2019intergenerational}, who compare mobility across regions in Canada and the United States; \citet{corak2020canadian}, who links regional mobility in Canada to local inequality; \citet{butikofer2022breaking}, who show that Norway's oil boom increased mobility in affected regions; \citet{acciari2022and}, who study direct income mobility across Italian regions; \citet{bell2023land}, who trace occupational mobility across areas in England and Wales across cohorts; \citet{dodin2024social}, who document regional variation in educational mobility in Germany; \citet{NybomStuhlerSMR}, who examine multigenerational mobility across Swedish municipalities; or \citet{Greboletal2025}, who link regional variation in Spanish educational mobility to differences in assortative mating.}, and surname-based measures may capture the relative ranking of mobility across regions (\citealt{guell2018correlating}).

However, there are theoretical reasons why surname-based and direct estimators might not yield the same regional rankings. In particular, two regions with identical parent-child correlations may nonetheless exhibit different surname-based estimates for two reasons: differences in surname distributions, and differences in how surname-based estimators weight persistent transmission pathways. The first source of divergence is mechanical and does not reflect actual differences in mobility. Because more common surnames exhibit higher persistence (Section \ref{subsec:namefreq}), regions with a greater concentration of such names will mechanically appear less mobile when using surname-based measures, even if underlying mobility is the same. \citet{guell2018correlating} highlight this issue and propose focusing on rare surnames in regional comparisons.

The second source, by contrast, does depend on the underlying intergenerational process. When parent-child correlations and surname distributions are similar, higher surname-based estimates indicate that persistent traits play a larger role in one region than in another, as surnames act as markers for such traits (Section \ref{subsec:name_pathways}). In other words, regions may exhibit similar short-run mobility but differ in how outcomes persist across more distant generations.\footnote{Evidence on the stability of regional rankings across different mobility measures has been limited. For Italy, \citet{guell2018correlating} and \citet{acciari2022and} find broadly similar regional patterns using surname-based and direct estimators, respectively. Using Swedish register data spanning three generations, \citet{NybomStuhlerSMR} show that regional rankings are similar for parent-child and grandparent-child correlations, even though the latter tend to exceed the square of the former in most municipalities.} For instance, if racial inequality is more pronounced in one region, and surnames vary with race, then surname-based estimates will capture this persistence even if parent-child correlations appear similar across regions.

To illustrate these arguments, we compare direct parent-child and surname-based estimates across geographic units of varying sizes within the United States, using the 1920 Census. For each surname–area cell (e.g., individuals named Smith in Wisconsin), we compute the average occupational score among working-age males, and assign these averages to children via their linked fathers. Using the same sample of males aged 30–40 in 1940 linked to their fathers, we then estimate the direct parent-child and surname-based measures separately within each geographic unit. We repeat this exercise at different levels of aggregation, starting from broad Census regions (Northeast, Midwest, South, and West), then moving to states, and finally to counties.\footnote{We use State Economic Areas (SEAs), which consist of single counties or groups of contiguous counties with similar economic characteristics \citep{ruggles2025ipums}.}
\begin{figure}[t]
  \centering
  \subfloat[Census and States]{\includegraphics[width=0.49\textwidth]{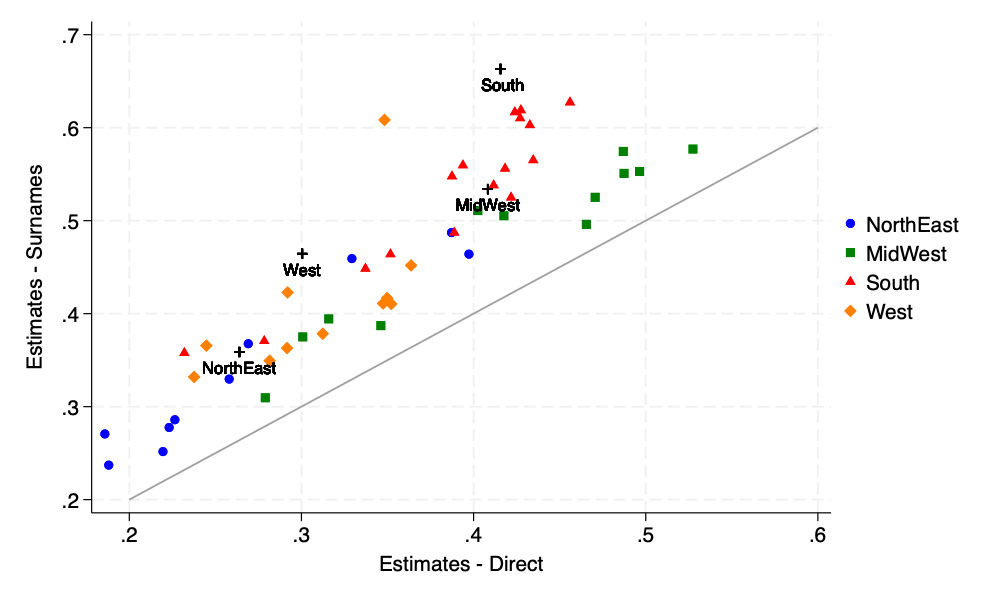}
  }\hfill
   \subfloat[Counties]{\includegraphics[width=0.49\textwidth]{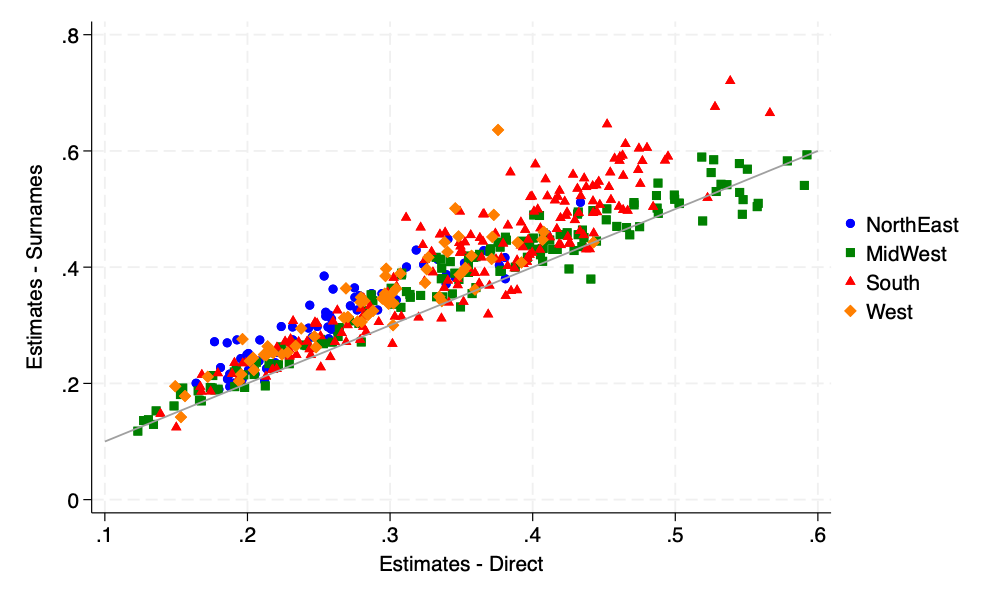}}
  \caption{Comparing Direct and Grouping Estimators Across U.S. Geographic Units}
  \label{geo_comp}
\floatfoot{\footnotesize \textit{Note:} This figure compares surname-based and direct estimates of intergenerational mobility across U.S. Census regions, states, and counties. For the surname-based estimator, we compute average log-occupational scores for each surname–area pair (e.g., all working-age males with a given surname within a given region or state) using the 1920 Census, and assign these values to children through their linked fathers. The estimate is then obtained by regressing the child’s log-occupational score on this surname–area average within each geographic unit. The direct estimate is obtained by regressing the child’s log-occupational score on the father’s log-occupational score within the same geographic unit. The sample consists of male workers aged 30–40 in the 1940 Census linked to their fathers.}
\end{figure}

Figure \ref{geo_comp} compares direct and surname-based estimates across geographic units. First, we confirm a key prediction of our model: surname-based estimators systematically exceed direct parent-child estimates, especially when considering large geographic areas such as Census regions or entire states (Panel A), and less so when considering individual counties (Panel B). This pattern is consistent with a greater reduction in idiosyncratic noise when averaging across more individuals in larger areas, and with surname-based estimators placing substantial weight on geographic differences. As we condition on increasingly smaller geographic units, the gaps between the two estimators decline, by progressively removing one of the main sources of their divergence.

Second, we evaluate the extent to which regional rankings are preserved. We find that regional rankings are broadly consistent regardless of whether mobility is measured directly or inferred from surnames. In line with earlier studies (\citealt{connor2020changing,tan2023different}), the Northeast emerges as the most mobile region in our early 20th century sample, while the South exhibits the lowest mobility. Midwestern states, especially non-industrial ones, show relatively high persistence in occupational status, a pattern that differs from more recent evidence (\citealt{chetty2014land}). 

Yet, even though rankings are broadly aligned, the gap between the two estimators varies substantially across regions. In particular, the gaps are larger in the U.S. South than in the Midwest. Although both regions display similar persistence using direct estimates, surname-based measures imply higher persistence in the South. This discrepancy may reflect differences in surname distributions or region-specific factors, such as racial inequality, that are more persistent and better captured by surnames. A similar, though less pronounced, pattern appears in the West.

Overall, although surname-based estimators identify a different parameter and tend to differ from direct measures, regional rankings are broadly similar. At the same time, the differences between them are informative, as they may reflect the influence of persistent local factors not captured by conventional measures.

\section{Multigenerational and Extended-Kin Estimates}\label{sec:multigenerational}

The indirect methods discussed in the previous sections are typically motivated by data \textit{limitations}: researchers lack data linking incomes for two generations and thus impute incomes using other characteristics, such as occupation, education, or even surnames. In contrast, another strand of research exploits the growing availability of data that link not only parents and children, but track broader family trees or multiple generations. Although the motivation differs, this literature shares commonalities with the instrumental variable (TSLS) and name-based approaches discussed earlier, both in implementation and interpretation. Through the lens of our transmission model, each approach produces differently weighted combinations of the underlying transmission mechanisms. While for TSLS and name-based estimators this reweighting arises as an unintended byproduct, it tends to be a more deliberate aspect of multigenerational and extended-kin studies. 

Recent multigenerational studies imply that socioeconomic inequalities persist across generations to a much greater extent than previously thought. In an early contribution, \cite{hodge1966occupational} cautioned that occupational mobility might not follow a simple first-order Markov process, in which children’s outcomes depend solely on those of their parents. However, researchers lacked data to study this question, and the scarce evidence that was available indicated that intergenerational correlations decayed quickly—consistent with the view of \citet{becker1986human} that  ``\textit{almost all earnings advantages and disadvantages of ancestors are wiped out in three generations}''.\footnote{Most prominently, \citet{warren1997social} analyzed three generations in Wisconsin and found that once parental socioeconomic characteristics were accounted for, grandparental occupational status had little predictive power for grandchildren’s outcomes. Similarly, \citet{erola2007social} reported that in Finland, the effect of grandparental social class on grandchildren’s class status was negligible when parental class was taken into account.} However, a pivotal shift in the economics literature occurred with \citet{lindahl2015long}, who integrated survey data from Sweden’s ``Malmö study'' with administrative records to examine earnings across three generations and educational attainment across four. Their findings revealed that multigenerational persistence was far stronger than what could be inferred by simply extrapolating two-generation regression estimates. Around the same time, other disciplines also adopted multigenerational perspectives, including sociology (\citealt{chan2013grandparents,hallsten2017grand, song2021multigenerational}), demography (\citealt{mare2011multigenerational}), and economic history (\citealt{dribe2016lasting}).

Since then, a growing body of research has shown that conventional models systematically underestimate the extent of long-term socioeconomic persistence (see \citealt{anderson2018grandparent, stuhler2024multigenerational}). If mobility followed an autoregressive AR(1) process, where each generation’s socioeconomic position depends only on that of their parents, the correlation between grandparent and grandchild outcomes (G1-G3) should roughly equal the product of the underlying parent-child correlations (G1-G2 and G2-G3). However, the evidence consistently shows that multigenerational correlations tend to be stronger than a naive iteration of the available parent-child evidence would suggest (the ``\textit{iterated regression fallacy}'', see \citealt{stuhler2012mobility}).\footnote{This pattern applies not only to income but also to other outcomes. For example, grandparental wealth remains a strong predictor of grandchildren’s wealth, even after controlling for parental wealth, in Sweden \citep{adermon2018intergenerational}, Denmark \citep{boserup2013intergenerational}, and the United States \citep{pfeffer2018generations}. While direct bequests that bypass the parent generation provide a plausible explanation, wealth-based advantages seem to take effect much earlier than such delayed transfers would suggest. Most studies focus on developed countries, with some recent exceptions (e.g., \citealt{celhay2025educational, CattanIndonesia}).}  Similarly, if mobility followed an AR(1) process, grandparental status should have little additional predictive power once parental characteristics are accounted for. In practice, however, most studies find that even after controlling for parental traits, grandparental and extended family socioeconomic status remain significant predictors of child outcomes (\citealt{anderson2018grandparent,solon2018we}).\footnote{Part of this pattern is due to the omission of the second parent (\citealt{Breen2018,braun2018transmission}). For example, \citet{engzell2020s} show that in Swedish data, grandparental status is not very predictive of child status once both maternal and paternal characteristics are accounted for. Measurement error in parental income may also inflate grandparent coefficients in multigenerational regressions (\citealt{ modalsli2024spillover}).}

To better understand the structural processes underlying intergenerational transmission, researchers increasingly incorporate additional empirical moments. Some studies extend the analysis beyond three generations—for example, \citet{hallsten2023shadow} examine persistence across up to seven generations in Sweden, \citet{modalsli2023multigenerational} considers five generations in Norway, and \citet{ward2025like} trace four generations in U.S. historical data. Other studies trace both vertical and horizontal links—while administrative registers rarely cover more than two or three generations, they are well suited for identifying lateral kinship links within extended families. By tracing relationships through siblings-in-law, cousins, and other distant kin, researchers can gain a more comprehensive picture of socioeconomic persistence. This dynastic or extended-family approach—long used in population genetics—is now increasingly applied in the social sciences (\citealt{adermon2021dynastic,collado2023estimating,clark2023inheritance,hallsten2023shadow,bingley2023origins,AdermonetalEthnicCapital}).

In this section, we review the multigenerational and extended kin literature through the lens of the ``multiplicity'' model described in Section \ref{sec:model}. The model can serve as a useful reduced-form framework for explaining key findings in this literature and for interpreting empirical moments beyond traditional parent-child correlations. Of course, the observation of high multigenerational correlations does not, in itself, point to any particular mechanism; different models, each with distinct implications, could account for the emerging evidence on multigenerational persistence (\citealt{stuhler2024multigenerational}). Nevertheless, our framework captures two practical features of intergenerational transmission that we expect to be important: that income is transmitted through multiple pathways, or mediators, and that different pathways exhibit varying rates of persistence.

\subsection{Multigenerational Correlations}

The recent literature has emphasized the importance of examining status correlations across multiple generations to better understand intergenerational transmission processes. 
We can again use the multiple latent factor model from Section \ref{sec:model} to understand why a comparison of inter- and multigenerational correlations can provide useful insights about both short- and long-term transmission.

The parent-child correlation in status $y$ as a function of our model is given by equation (\ref{eq:corr2g}). The corresponding model-implied formula for the \textit{grandparent}-child correlation is
\begin{align}\label{eq:corr3g}
    \beta_{gp}=\frac{Cov(y_{it};y_{it-2})}{V(y_{it-2})}=\frac{\sum_{j=0}^{J}\lambda_{j}^{2}\omega_{j}}{\sum_{j=0}^{J}\omega_{j}+\omega_{u}}
\end{align}
where 
\begin{align}\label{eq:corr3gweights}
    \omega_{j}=\rho_{j}^{2}V(X_{it-2}^{j})+\rho_{j}\sum_{k\neq j}^{J}\rho_{k}Cov(X_{it-2}^{j};X_{it-2}^{k})
\end{align}
As for the parent-child correlation, the grandparent-child correlation corresponds to a weighted average of multiple transmission pathways, but the contribution of each characteristic $j$ is now proportional to the \textit{square} of its transferability $\lambda_j$. As a consequence, the relative importance of different characteristics for the overall correlation shifts: characteristics that are less relevant in the short term (low $\omega_{j}$) may explain a greater share of the grandparent-child correlation if they are highly persistent (high $\lambda_{j}$). Conversely, characteristics that are highly relevant in the short term (high $\omega_{j}$) but not very transmissible (low $\lambda_{j}$) will play a smaller role in long-term correlations.

Thus, parent-child and grandparent-child correlations capture differently weighted combinations of the underlying pathways of intergenerational transmission. We draw two conclusions from this observation. First, neither set of weights is inherently superior. Although the literature has traditionally focused on parent-child correlations, it is important to note that they offer only limited insight into mobility processes. Most obviously, they will provide little insight into transmission patterns across multiple generations.\footnote{In particular, the model-implied grandparent-child correlation is generally not equal to the square of the parent-child correlation, even if all moments are in steady state. If $\omega_u>0$, or if the transferability $\lambda_j$ varies across characteristics—as is intuitive and also suggested by the surname-based evidence from Section \ref{sec:names}—status decays at less than a geometric rate.} Second, by comparing mobility measures across close and more distant kin, researchers can gain a deeper understanding of intergenerational transmission processes, as we illustrate next.

\subsection{Instrumenting Close with Distant Kin}

While any given mobility measure provides only a limited understanding of intergenerational processes, a comparison of different measures can yield additional insights. From the perspective of the latent transmission model, an especially promising strategy is to compare correlations between close and more distant kin—possibly using instrumental variable strategies that share some similarity with the IV strategies discussed in previous sections.

\textit{The simple case with a single latent factor.} To illustrate the basic idea, consider the simple latent factor model with a single latent characteristic $X_{it}^j$, or alternatively, multiple characteristics with the same transferability $\lambda_j=\lambda$ (see restriction (\ref{rest})). As noted in \cite{braun2018transmission}, the transferability of the latent advantage(s) can then be identified by using the grandparental outcome as an instrument for the parental outcome,\footnote{To account for changes in the variance of the outcomes across generations, \cite{braun2018transmission} replace covariances with correlations—which is equivalent to standardizing the outcome variable prior to applying the instrumental variable method.} 
\begin{equation}\label{eq:LFM_IV}
  \beta_{IV}\equiv\frac{Cov(y_{it};y_{it-2})}{Cov(y_{it-1};y_{it-2})}=\frac{\lambda^{2}\sum_{j=0}^{J}\omega_{j}}{\lambda\sum_{j=0}^{J}\omega_{j}}=\lambda
\end{equation}
That is, even if we cannot directly observe the latent advantages $X_{it}^j$, comparing close and distant kin allows researchers to identify its transferability $\lambda$ (and to eliminate the influence of measurement error). This example illustrates the more general insight that the covariances between an outcome and its lags can be a powerful source for identifying latent variable models (\citealt{aigner1984latent}).\footnote{In practice, the identifiability of intergenerational transmission models also depends on steady-state assumptions (\citealt{NybomStuhler2019}) and the available data structure. \citet{FamilyCookbook} provide a systematic guide to the identification of intergenerational models, showing how different combinations of kinship moments and assumptions—including steady-state restrictions and assortative assumptions—determine which parameters can be recovered.}

This particular way to compare parent-child and multigenerational correlations has been adopted in several recent studies. \cite{braun2018transmission} analyze multigenerational persistence in education and occupation in German samples, estimating a parent-child correlation in ‘latent’ advantages of about 0.6—nearly 50\% higher than the observed correlation in years of schooling. Using harmonized survey data, \cite{neidhofer2019dynastic} find slightly higher latent persistence (around 0.7) for Germany and somewhat lower rates for the U.S. and UK, while \cite{colagrossi2020like} report an average latent educational persistence of 0.66 across 28 European countries. A consistent finding across these studies is that latent persistence substantially exceeds the conventional parent-child correlation in observed outcomes.\footnote{Other studies interpreting multigenerational data through the lens of a latent factor model include \cite{lundberg2020does, barone2021intergenerational, benhabib2022heterogeneous, hallsten2023shadow, colagrossi2023intergenerational, li2023multi,belloc2024multigenerational}, and \cite{croix2024nepotism}. For example, \cite{belloc2024multigenerational} show that such a model could account for the pattern of wealth persistence across many generations in historical data from Italy. \cite{croix2024nepotism} consider an extended latent factor model to study nepotism and human capital transmission in academic dynasties.}

\textit{The general case with multiplicity.} The result in equation (\ref{eq:LFM_IV}) hinged on the assumption that the transferability does not vary across characteristics. In the more general case of a model with multiple transmission pathways, the IV estimator identifies
\begin{align}\label{eq:MLFM_IV}
    \beta_{IV}\equiv\frac{Cov(y_{it};y_{it-2})}{Cov(y_{it-1};y_{it-2})}=\frac{\sum_{j=0}^{J}\lambda_{j}^{2}\omega_{j}}{\sum_{j=0}^{J}\lambda_{j}\omega_{j}}
\end{align}
where $\omega_{j}$ is defined in (\ref{eq:corr3gweights}). In this more general case, the IV approach still identifies a weighted combination of the latent transferability parameters $\lambda_j$. However, the estimator tends to weight more heavily those characteristics that are more persistent. To see this, note that expression (\ref{eq:MLFM_IV}) for the multigenerational estimator differs from the corresponding expression (\ref{eq:corr2g}) for the direct estimator in that the contribution of each characteristic $j$ is scaled by its transferability $\lambda_j$.\footnote{This also implies that the term $\omega_u$ in the denominator of expression (\ref{eq:corr2g}) cancels out, as the noise $u$ is not transferable across generations ($\lambda_u=0$).} As is intuitive, characteristics shared between grandparents and grandchildren will tend to be more persistent than those shared between parents and children. 

There are thus both important parallels and differences between this IV estimator and the TSLS and name-based estimators reviewed in earlier sections. One distinction is that in multigenerational and extended-kin studies, researchers typically observe parental outcomes. Thus, rather than a workaround for missing data, instrumentation is used as a deliberate strategy to gain deeper insights into the transmission process.  Nevertheless, their interpretation is subject to similar considerations. Our key argument regarding all indirect estimators is that they should be interpreted as differently weighted combinations of the underlying transmission mechanisms. In the TSLS and name-based approaches, this reweighting emerges as an unintended byproduct; in multigenerational and extended-kin studies, it tends to be more directly tied to the rationale for adopting the approach in the first place.

\textit{More kin.} More generally, transmission parameters could be recovered through a variety of family relationships. A simple approach is to use the ratio between grandparent and parent-child correlations, as shown above. However, this logic can be extended beyond grandparents to encompass any distant relative, including \textit{affine} kin (relatives by marriage). For instance, we can also leverage relationships involving spouses and parents-in-law. By defining the spousal correlation in trait $j$ as $\gamma_{j}$ and relabeling the vertical transmission parameters $\lambda_{j}=\tilde{\lambda}_{j}\frac{(1+\gamma_{j})}{2}$ to account for assortative mating, we can recover vertical transmission estimates through the ratio of covariances between the individual and their parent-in-law and between the individual and their spouse:
\begin{align}\label{eq:beta_parentinlaw_spouse}
\frac{Cov(y_{i,t};y_{pil,t-1})}{Cov(y_{i,t};y_{sp,t})}&=\frac{\sum_{j=1}^{J}\lambda_{j}\omega_{j}}{\sum_{j=1}^{J}\omega_{j}}
\end{align}
where 
\begin{align*}
\omega_{j}&=\rho_{j}^{2}V(x_{i,t-1}^{j})\gamma_{j}
\end{align*}
Or, defining the sibling correlation in trait $j$, denoted as $r_{j}$,  a parallel strategy can be applied using cousins and uncles/aunts:
\begin{align}\label{eq:beta_cousin_uncles}
\frac{Cov(y_{i,t};y_{cousin,t})}{Cov(y_{i,t};y_{uncle,t-1})}&=\frac{\sum_{j=1}^{J}\lambda_{j}\omega_{j}}{\sum_{j=1}^{J}\omega_{j}}
\end{align}
where
\begin{align*}
\omega_{j}&=\rho_{j}^{2}V(x_{i,t-1}^{j})\lambda_{j}r_{j}
\end{align*}

\begin{figure}[t]
    \centering
    \includegraphics[width=0.7\linewidth]{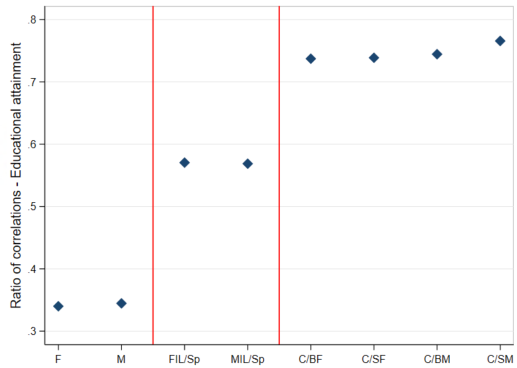}
    \caption{Instrumenting Close With Distant Kin}
    \label{fig:iv_p}
    \floatfoot{\footnotesize \textit{Note:} The figure compares father-child and mother-child correlations in years of schooling (``Parents''), the ratio of the parent-in-law correlation to the spouse correlation (``Parents in law'', eq. (\ref{eq:beta_parentinlaw_spouse})), and the ratio of the uncle/aunt--nephew/niece correlation to the cousin correlation (``Uncle/Aunt'', eq. (\ref{eq:beta_cousin_uncles})). On the horizontal axis, $F$ and $M$ denote father and mother. $FIL/Sp$ and $MIL/Sp$ denote the ratio of the father-in-law and mother-in-law correlation to the spouse correlation, respectively. $C/BF$, $C/SF$, $C/BM$, and $C/SM$ denote the ratio of the cousin correlation to the correlation with the father's brother, father's sister, mother's brother, and mother's sister, respectively. Source: Swedish register data (2020 extract), individuals born before 1990.}
\end{figure}

In general, as we rely on increasingly distant relatives, the estimation should become progressively more weighted towards more persistent traits. We can provide empirical support for this reasoning using Swedish register data. As shown in Figure \ref{fig:iv_p}, moving from direct parent-child correlations in years of schooling (shown on the left) to alternative estimators based on the more distant family links considered in equations (\ref{eq:beta_parentinlaw_spouse}) and (\ref{eq:beta_cousin_uncles}) results in progressively higher estimates of the underlying vertical transmission parameter.

These arguments are again closely related to our discussion of name-based estimators in Section \ref{sec:names}: by considering more distant relatives—either explicitly in the extended-kin approach, or implicitly by focusing on larger name groups in name-based estimators—researchers place more weight on the more persistent factors that shape long-run transmission. Whether this is a flaw or a feature depends on one's research objective. Ultimately, the observation that different estimators identify differently weighted combinations of transmission pathways leads to the argument that researchers should systematically compare multiple mobility measures across both close and distant kin.

\subsection{Dynastic and Extended-Kin Studies}

The previous section illustrated the potential usefulness of instrumental variable approaches that instrument close with more distant kin. However, it also illustrated that only in the simplest models, such estimators map directly into structural parameters; more generally, they identify a weighted combination of different parameters, as in the case of the multiple latent factor model in expression (\ref{eq:MLFM_IV}). A useful generalization of the IV approach is therefore to compare \textit{many} kinship moments, and to use the observed patterns to fit more general models of intergenerational transmission. 

To do so, researchers go beyond parent-child links and consider broader kinship networks. While administrative data rarely span more than two or three generations, they are particularly valuable for identifying lateral \textit{horizontal} links across extended families. By tracing relationships through siblings-in-law, cousins, and other distant relatives, researchers can construct a richer and more nuanced picture of socioeconomic persistence. The value of systematic comparisons across close and distant kin has long been recognized in population and behavioral genetics (“correlation between relatives”; see \citealt{crowkimura1970, cavalli1981cultural, reynolds2000mixed, kemper2021phenotypic}). In the social sciences, however, this approach has only recently begun to gain traction \citep{adermon2021dynastic, collado2023estimating, hallsten2023shadow, clark2023inheritance, bingley2023origins, colagrossi2024heterogenous,AdermonetalEthnicCapital}.

A pioneering study in this area is \citet{adermon2021dynastic}, who show that ``dynastic'' human capital—the socioeconomic status of parents' siblings, cousins, and their spouses—significantly predicts children’s socioeconomic outcomes. Consistent with findings in the multigenerational literature, even after controlling for parental characteristics, the socioeconomic status of parents' siblings, cousins, and their spouses remains a strong predictor of child outcomes. Their results imply that conventional parent-child regressions underestimate true long-run persistence by at least one-third. Most of this added persistence can be captured by considering the extended family in the parental generation, without including data from the grandparent or great-grandparent generations.\footnote{\citet{AdermonetalEthnicCapital} show that extended-family influences can also account for much of what previous research attributed to ``ethnic capital''—the independent effect of origin-group status on children's outcomes.}

\citet{collado2023estimating} interpret extended kinship data from Sweden through the lens of a generalized latent factor model, showing that a unified model can provide a good fit—both in and out of sample—to a set of 141 distinct kinship moments. They find strong intergenerational persistence in the latent determinants of education and income. Moreover, their findings imply a striking degree of sorting, as kinship correlations decay only slowly, by around one quarter, with each additional spousal link. To explain such slow decay, spouses must share advantages to a far greater extent than indicated by conventional measures of sorting, such as the spousal correlation in years of schooling.

\citet{clark2023inheritance} uses genealogical data from England spanning 1600 to 2022 to trace status correlations across relatives as distant as fourth cousins, finding that correlations decline only slowly with genetic distance, consistent with strong assortative mating.\footnote{\citet{clarkcummins2022assortative} estimate the underlying spousal correlation in occupational status in England from 1754 to 2021, finding a value of approximately 0.8—far higher than the conventionally measured correlation of 0.39—which they attribute to measurement error in observed status indicators.} An interesting question is whether extended-kin data can discriminate between competing models of intergenerational transmission, as structurally different models may generate similar predictions for observable correlations. For instance, \citet{collado2023estimating} find that a standard additive genetic model with phenotypic assortative mating cannot fit the Swedish data well, while \citet{clark2023inheritance} argues that such a model — and assuming that mating occurs on a broader, partially observed phenotype — can account for the kinship patterns in English data spanning four centuries.

An illustration of how incorporating extended kin can refine our understanding of underlying transmission mechanisms comes from \citet{bingley2023origins}, who reassess the assumptions underlying the widely used ACE model in behavioral genetics. This model decomposes variance in outcomes such as education, earnings, or wealth into additive genetic (A), shared environmental (C), and non-shared environmental (E) components. By incorporating extended kin,
the authors can relax some of the model’s key assumptions, which significantly reduces the estimated genetic component (A), while increasing the estimated influence of shared environmental factors—challenging earlier twin-based estimates of high heritability in socioeconomic outcomes.

\section{Concluding Remarks}

In this chapter, we interpreted popular ``indirect'' estimators of intergenerational mobility through a common framework that allows for multiple transmission pathways with heterogeneous transmission rates. We highlighted important parallels across classic instrumental-variable, name-based, and extended-kin approaches, arguing that each can be viewed as estimating a differently weighted combination of the mechanisms through which socioeconomic status is transmitted. Notably, name-based estimators give greater weight to more persistent pathways than do conventional parent-child correlations. While this might constitute a drawback for some purposes, it also makes name-based estimators informative about the long-run dynamics of intergenerational transmission. 

An implication is that researchers can learn more by comparing multiple mobility measures instead of relying on a single measure. In particular, conventional parent-child correlations may provide only limited insight into the strength of the underlying transmission processes, the persistence of broader socioeconomic advantages, or mobility beyond two generations. Instead, a promising direction is to systematically compare correlations across both close and distant kin, as such comparisons can reveal the strength of latent transmission processes and help quantify how socioeconomic inequalities persist over both the short and long run.

\newpage
\bibliographystyle{chicago}
\bibliography{handbook_source}

\newpage
\appendix
\section{Appendix}
\renewcommand{\thefigure}{A.\arabic{figure}}
\renewcommand{\thetable}{A.\arabic{table}}

\setcounter{figure}{0}
\setcounter{table}{0}

\subsection{Data}

\subsubsection{Swedish register data}\label{sec:swe_data}

To provide empirical illustrations of IV estimators in Section \ref{sec:IV}, we use data from Swedish administrative registers. We link fathers born 1945-1960 to children using birth registers. We have access to earnings data from 1968-2020, and compute long-run earnings of fathers as the mean of annual earnings between ages 40 and 50, and for children as the mean between ages 30 and 40. We bottom-code all non-missing annual observations at 10,000 SEK (roughly 1,000 USD) to decrease the influence of very low incomes on IGE estimates. We drop fathers with fewer than seven annual earnings observations and children with fewer than two earnings observations. 

We then construct residualized log earnings and earnings ranks for both fathers and children. For fathers, we take logs of the average and residualize with respect to (fathers') year-of-birth dummies. Alternatively, we rank earnings by year of birth. For children, we residualize or rank incomes by child year of birth and gender.

We add to the data information on years of schooling (transformed from highest level of attainment) from the Education register, and information on fathers' occupation, municipality of residence, and country of birth from the 1990 census. For the children, we add the corresponding information from 2015, using population registers. For a subsample, we also add scores on cognitive and non-cognitive skill tests from the military enlistment, based on separate tests taken by men born from 1951 and until the mid 1980s. For these cohorts, the draft was mandatory for nearly all men (exemptions were given primarily due to disabilities or other health reasons) and featured a relatively standard but extensive written test of cognitive skill, similar to the Armed Forces Qualification Test. Draftees were also scored on their “non-cognitive”
skill, e.g., their ability to interact with others and their leadership abilities, based on a roughly half-hour long semi-structured interview with a certified psychologist. Both tests were scored on so-called Standard nine scales, stretching between one and nine with mean five and standard deviation two. We standardize each test score to have mean zero and standard deviation one by year of enlistment.

For the illustration of extended-kin correlations in Section \ref{sec:multigenerational}, specifically Figure \ref{fig:iv_p}, we use data from the Swedish population registers for the year 2020, restricting the sample to individuals aged 30 and above. We link individuals to their parents, siblings, and cousins using information from the birth registers. For each individual, we then again merge information on years of schooling as derived from the Education register.

\subsubsection{U.S. Census data}\label{sec:us_data}
In Section \ref{sec:names}, we draw on data from the 1920 and 1940 U.S. Censuses, accessed through IPUMS. To compute surname-based averages, used as regressors in our surname-level estimators, we include the full working-age male population (ages 21 to 64) from the 1920 Census. Our main sample for intergenerational analysis consists of male children who were between 10 and 20 years old in 1920 and who could be successfully linked to their 1940 Census records, where they appear aged 30 to 40. We use links provided by the CensusTree project (\citealt{buckles2023breakthroughs}), achieving a match rate exceeding 50\%\footnote{During the linking process, some individuals are matched with inconsistent ages across censuses. We exclude cases where the implied age difference exceeds five years, while retaining smaller discrepancies to account for potential reporting errors in age.}. We further restrict the sample to individuals for whom both individual-level and surname-level measures of paternal outcomes are observed. Surname-level averages are computed using the full working-age male population in the 1920 Census and are assigned to children through their linked fathers. This assignment through the father’s observed surname also avoids measurement error from misspelling.

Because the 1920 Census lacks direct income data, we use log-occupational scores (``occscore'') as a proxy for socioeconomic status. This score is a constructed variable that assigns occupational income scores to each occupation, based on the median income in each occupation in 1950. In addition to occupational scores, our dataset includes several covariates: educational attainment from the 1940 Census, state and county of residence in 1940, birthplace, and an indicator for urban versus rural status.
 
In addition to the baseline sample, we construct two subsamples to assess the sensitivity of surname-based estimators to sampling variation and overlap. First, we draw a random 5\% subsample of linked father–son pairs, i.e. males aged 30–40 in 1940 who can be linked to their fathers observed in 1920. Within this subsample, surname averages are constructed using only the fathers observed in the same sample, implying full overlap between individual outcomes and surname-level averages. Second, we draw a separate random 5\% sample of working-age males from the 1920 Census and use this sample to construct surname averages. These averages are then merged to the 5\% subsample of children described above. Because the two samples are drawn independently, this procedure implies limited overlap between the fathers of the estimation sample and the individuals used to construct surname averages. These alternative constructions allow us to isolate the effects of sampling noise and overlap on the properties of surname-based estimators, as discussed in Section \ref{subsec:names_TSIV}.

To study regional variation in intergenerational mobility, we construct surname averages at different levels of geographic aggregation using the male working-age population in 1920. Specifically, we interact surnames with geographic identifiers in the 1920 Census and compute average outcomes within each surname–area cell (e.g., individuals with a given surname within a state or county). These averages are then assigned to children through their linked fathers.

\subsection{Additional Derivations}

\subsubsection{Derivation of equation (\ref{eq:varX_name_dist})}
\label{proof_name}
In this section, we derive equation (\ref{eq:varX_name_dist}). For the  sake of simplicity we assume that each characteristic is centered around zero.
From the model equation on transmission for characteristic $k$:
\begin{align*}
E[X_{ist}^{j}|sur=s]=\lambda_{j}E[X_{ist-1}^{j}|sur=s]+\underbrace{E[\epsilon_{ist}^{j}|sur=s]}_{=0}
\end{align*}
by independence of the errors $\epsilon_{ist}^{j}$ in the relation between child and parent characteristics, $E[\epsilon_{ist}^{j}|sur=s]=E[\epsilon_{ist}^{j}]=0$.
We iterate this procedure until we encounter the common ancestor for surname $s$ in generation $\tau_{s}$, whose value for characteristic $j$ is taken as given,
 \begin{align*}
     E[X_{is\tau_{s}}^{j}|sur=s]=X_{is\tau_{s}}^{j}
 \end{align*}
such that
\begin{align*}
    E[X_{ist}^{j}|sur=s]=\lambda_{j}^{t-\tau_{s}}X_{is\tau_{s}}^{j}
\end{align*}
Now, we take the variance of this term
\begin{align*}
    V(E[X_{ist}^{j}|sur])&=V(\lambda_{j}^{t-\tau_{s}}X_{is\tau_{s}}^{j})\\
    &=E[(\lambda_{j}^{t-\tau_{s}}X_{is\tau_{s}}^{j})^{2}]-E[\lambda_{j}^{t-\tau_{s}}X_{is\tau_{s}}^{j}]^{2}
\end{align*}
We assume independence of $X_{is\tau_{s}}^{j}$ from $\tau_{s}$, such that the size of a surname group is not correlated with the characteristics of the common ancestor
\begin{align*}
    E[(\lambda_{j}^{t-\tau_{s}}X_{is\tau_{s}}^{j})^{2}]-E[\lambda_{j}^{t-\tau_{s}}X_{is\tau_{s}}^{j}]^{2}&=E[\lambda_{j}^{2(t-\tau_{s})}]V(X_{is\tau_{s}}^{j})-E[\lambda_{j}^{t-\tau_{s}}]^{2}\underbrace{E[X_{is\tau_{s}}^{j}]^{2}}_{=0}\\
    &=E[\lambda_{j}^{2(t-\tau_{s})}]V(X_{is\tau_{s}}^{j})
\end{align*}
Similarly, the corresponding covariances can be derived as 
\begin{align*}  Cov(E[X_{ist}^{j}|sur],E[X_{ist}^{k}|sur])=Cov(\lambda_{j}^{t-\tau_{s}}X_{is\tau_{s}}^{j},\lambda_{k}^{t-\tau_{s}}X_{is\tau_{s}}^{k})=E[\lambda_{j}^{t-\tau_{s}}\lambda_{k}^{t-\tau_{s}}]Cov(X_{is\tau_{s}}^{j},X_{is\tau_{s}}^{k})
\end{align*}

\subsubsection{Alternative Expression for Bias in TSLS Estimator}
\label{rel_lit}
In this section, we provide intuition on how our framework relates to standard results in the literature. For simplicity, we assume that the latent components $X^j$ are mutually independent.

Consider the linear projection of child income on parental income, where $\beta_{IGE}$ denotes the true intergenerational elasticity:
\begin{align*}
y_{it} &= \beta_{IGE} y_{it-1} + \zeta_{it}.
\end{align*}
In a two-stage least squares (TSLS) setting, assuming that the fitted values converge to the true conditional expectation $\hat{y}_{it-1} = E[y_{it-1} \mid Z_{it-1}]$, the TSLS estimator can then be expressed as:
\begin{align*}
\beta_{\text{TSLS}} 
&= \frac{\operatorname{Cov}(y_{it}, \hat{y}_{it-1})}{\operatorname{Var}(\hat{y}_{it-1})} \\
&= \beta_{IGE} + \frac{\operatorname{Cov}(\zeta_{it}, \hat{y}_{it-1})}{\operatorname{Var}(\hat{y}_{it-1})}.
\end{align*}
As is standard, the bias depends on the sign of the covariance term. Our framework helps clarify the conditions under which this bias is positive or negative—that is, when TSLS over- or underestimates the true IGE.

We can express the unexplained component of child income as:
\begin{align*}
\zeta_{it} 
&= y_{it} - \beta_{IGE} y_{it-1} \\
&= \sum_{j \in J} \rho_j (\lambda_j X^{j}_{t-1} + \epsilon^{j}_{it}) 
   - \beta_{IGE} \sum_{j \in J} \rho_j X^{j}_{t-1} \\
&= \sum_{j \in J} \rho_j \left[(\lambda_j - \beta_{IGE}) X^{j}_{t-1} + \epsilon^{j}_{it}\right].
\end{align*}
Using this decomposition,\footnote{This step additionally assumes that the projected components are orthogonal, i.e. $\operatorname{Cov}(E[X^j_{t-1}\mid Z_{it-1}], E[X^k_{t-1}\mid Z_{it-1}])=0$ for $j\neq k$.}
the covariance term can be written as:
\begin{align*}
\operatorname{Cov}(\zeta_{it}, \hat{y}_{it-1}) 
= \sum_{j \in J} \rho_j^2 (\lambda_j - \beta_{IGE}) 
  \operatorname{Var}\!\left(E[X^{j}_{t-1} \mid Z_{it-1}]\right).
\end{align*}
This expression directly determines the sign of the TSLS bias. If the conditional variance $\operatorname{Var}(E[X^{j}_{t-1} \mid Z_{it-1}])$ equals the true variance $\operatorname{Var}(X^{j}_{t-1})$—that is, if the instrument perfectly replicates the composition of the latent traits underlying the IGE—the covariance term is zero, implying no bias.

When the conditional variance is larger for traits with $\lambda_j > \beta_{IGE}$, the covariance becomes positive, generating an \textit{upward bias} in the TSLS estimate. Conversely, when the conditional variance is larger for traits with $\lambda_j < \beta_{IGE}$, the covariance is negative, leading to a \textit{downward bias}.

\subsection{Additional Figures and Tables}

\subsubsection{Swedish Data (Section 3)}

\begin{table}[H]
\centering
\begin{center}
\centering{}
{ 
\global\long\def\sym#1{\ifmmode^{#1}\else$^{#1}$\fi}%
 \footnotesize
\begin{tabular}{lccccccccc}
\toprule 
 & \multicolumn{1}{c}{$y_{t}$} & \multicolumn{1}{c}{$x\hat{\beta}_{t}$} & \multicolumn{1}{c}{$x\hat{\beta}_{t}$} & \multicolumn{1}{c}{$x\hat{\beta}_{t}$} & \multicolumn{1}{c}{$x\hat{\beta}_{t}$} & \multicolumn{1}{c}{$x\hat{\beta}_{t}$} & \multicolumn{1}{c}{$x\hat{\beta}_{t}$} & \multicolumn{1}{c}{$x\hat{\beta}_{t}$} & \multicolumn{1}{c}{}\tabularnewline
$X$-var.: & \multicolumn{1}{c}{} & \multicolumn{1}{c}{Educ.} & \multicolumn{1}{c}{Occup.} & \multicolumn{1}{c}{Region} & \multicolumn{1}{c}{Birth ctry.} & \multicolumn{1}{c}{Cogn.} & \multicolumn{1}{c}{Noncogn.} & \multicolumn{1}{c}{All $X$} & \multicolumn{1}{c}{Res.}\tabularnewline
 & (1) & (2) & (3) & (4) & (5) & (6) & (7) & (8) & (9)\tabularnewline
\hline
 \addlinespace[1ex]
\multicolumn{10}{c}{\underline{Panel A: Men}} \tabularnewline
 \addlinespace[1ex]
$y_{t-1}$ & 0.199 &  &  &  &  &  &  &  & \tabularnewline
 & (0.002) &  &  &  &  &  &  &  & \tabularnewline
$x\hat{\beta}_{t-1}$ &  & 0.227 & 0.073 & 0.574 & 0.229 &  &  & 0.205 & \tabularnewline
 &  & (0.001) & (0.001) & (0.002) & (0.001) &  &  & (0.001) & \tabularnewline
Residual &  &  &  &  &  &  &  &  & 0.111\tabularnewline
 &  &  &  &  &  &  &  &  & (0.002)\tabularnewline
$R^{2}_{\textit{first}}$ & 0.037 & 0.123 & 0.013 & 0.222 & 0.333 &  &  & 0.083 & 0.009\tabularnewline
$N$ & 395,424 & 395,424 & 395,424 & 395,424 & 395,424 &  &  & 395,424 & 395,424\tabularnewline
 \addlinespace[1.5ex]
\multicolumn{10}{c}{\underline{Panel B: Men, IQ sample}} \tabularnewline
 \addlinespace[1ex]
$y_{t-1}$ & 0.172 &  &  &  &  &  &  &  & \tabularnewline
 & (0.003) &  &  &  &  &  &  &  & \tabularnewline
$x\hat{\beta}_{t-1}$ &  & 0.191 & 0.064 & 0.602 & 0.013 & 0.227 & 0.223 & 0.210 & \tabularnewline
 &  & (0.002) & (0.002) & (0.003) & (0.001) & (0.002) & (0.004) & (0.003) & \tabularnewline
Residual &  &  &  &  &  &  &  &  & 0.086\tabularnewline
 &  &  &  &  &  &  &  &  & (0.004)\tabularnewline
$R^{2}_{\textit{first}}$ & 0.032 & 0.096 & 0.009 & 0.279 & 0.002 & 0.114 & 0.038 & 0.073 & 0.007\tabularnewline
$N$ & 83,255 & 83,255 & 83,255 & 83,255 & 83,255 & 83,255 & 83,255 & 83,255 & 83,255\tabularnewline
\bottomrule
\end{tabular}}
\end{center}
\caption{Intergenerational Persistence of Different Characteristics}
\label{igps}
\vspace{-0.6cm}
\floatfoot{\footnotesize \textit{Note:} OLS regressions in Swedish administrative registers.  Column (1) shows the ``direct'' OLS regression of child (sons) on father's log earnings. Columns (2)-(8) show estimates using fitted values from differents sets of characteristics for father's and children's log earnings: in col. (2) dummies for education level, in (3) 3-digit occupation dummies, in (4) region of residence dummies, in (5) dummies for country of birth, and in (6) and (7) measures of standardized cognitive and noncognitive skill from the military draft. All specifications are reestimated using the subsample with available draft data (``IQ sample''). Column (8) uses the fitted values from regressions including all characteristics listed in (2)-(7). Column (9) shows estimates from using the residuals from the joint regressions underlying col. (8). See Appendix \ref{sec:swe_data} for details.}
\end{table}

\subsubsection{U.S. Census Data (Section 4)}
\begin{table}[H]
 \centering
    \resizebox{0.8\textwidth}{!}{%
\begin{tabular}{lrrrrrr}
\toprule
 & \multicolumn{5}{c}{Quintiles of Surname Frequency} & \multicolumn{1}{c}{Total} \\
\cmidrule(lr){2-6}
 & \multicolumn{1}{c}{1} & \multicolumn{1}{c}{2} & \multicolumn{1}{c}{3} & \multicolumn{1}{c}{4} & \multicolumn{1}{c}{5} & \\
\midrule
Child Occupational Score & 3.168 & 3.146 & 3.140 & 3.139 & 3.106 & 3.140 \\
Years of Schooling & 9.437 & 9.525 & 9.556 & 9.554 & 9.295 & 9.473 \\[3pt]

\textit{Race} &  &  &  &  &  &  \\
\hspace{1em}White Non-Hispanic & (97.4\%) & (96.1\%) & (94.9\%) & (93.7\%) & (90.4\%) & (94.5\%) \\
\hspace{1em}Other & (2.6\%) & (3.9\%) & (5.1\%) & (6.3\%) & (9.6\%) & (5.5\%) \\[3pt]

\textit{Census Division} &  &  &  &  &  &  \\
\hspace{1em}NorthEast & (32.9\%) & (27.6\%) & (24.9\%) & (24.3\%) & (20.0\%) & (26.0\%) \\
\hspace{1em}MidWest & (41.9\%) & (37.1\%) & (33.2\%) & (31.2\%) & (30.7\%) & (34.8\%) \\
\hspace{1em}South & (15.6\%) & (25.0\%) & (31.1\%) & (33.4\%) & (38.2\%) & (28.7\%) \\
\hspace{1em}West & (9.5\%)  & (10.3\%) & (10.8\%) & (11.1\%) & (11.0\%) & (10.5\%) \\[3pt]

\textit{Urban/Rural Status} &  &  &  &  &  &  \\
\hspace{1em}Rural & (39.0\%) & (45.9\%) & (48.2\%) & (48.1\%) & (51.3\%) & (46.5\%) \\
\hspace{1em}Urban & (61.0\%) & (54.1\%) & (51.8\%) & (51.9\%) & (48.7\%) & (53.5\%) \\[3pt]

Family Size (1920) & 4.586 & 4.414 & 4.349 & 4.330 & 4.360 & 4.408 \\[3pt]

N & 899{,}350 & 890{,}930 & 894{,}589 & 896{,}679 & 893{,}058 & 4{,}474{,}606 \\
\bottomrule
\end{tabular}
}
\caption{Summary Statistics by Quintiles of Surname Frequency}
\label{tab:surname_freq_summary}
\floatfoot{\footnotesize \textit{Note:} This table reports summary statistics for the U.S. sample described in Appendix~\ref{sec:us_data}, by quintiles of surname frequency. For continuous variables, the table displays mean values. For categorical variables, it reports the share of individuals belonging to each category. Family size is measured as the number of co-resident children in the household in 1920.}
\end{table}

\end{document}